\documentclass[journal,twocolumn]{IEEEtran}
\ifCLASSINFOpdf
   \usepackage[pdftex]{graphicx}
\else
\fi
\usepackage{amsmath}
\usepackage{amssymb}
\usepackage{algpseudocode}
\usepackage{url}

\usepackage{xcolor}

\begin{document}
%
\title{Robust Real-Time Delay Predictions in\\a Network of High-Frequency Urban Buses}
%
%
%

\author{Hector~Rodriguez-Deniz and Mattias~Villani
\thanks{H. Rodriguez-Deniz is at the Division of Statistics and Machine Learning (STIMA), Department of Computer and Information Science (IDA), Link\"oping University, Link\"oping, Sweden (e-mail: hector.rodriguez@liu.se).}
\thanks{M. Villani is also at STIMA (Link\"oping University), and at the Department of Statistics, Stockholm University, Stockholm, Sweden (e-mail: mattias.villani@stat.su.se).}
}

%
%

\markboth{PREPRINT SUBMITTED TO IEEE}%
{MANUSCRIPT PREPRINT}
%



\maketitle

\begin{abstract}
Providing transport users and operators with accurate forecasts on travel times is challenging due to a highly stochastic traffic environment. Public transport users are particularly sensitive to unexpected waiting times, which negatively affect their perception on the system’s reliability. In this paper we develop a robust model for real-time bus travel time prediction that departs from Gaussian assumptions by using Student-$t$ errors. The proposed approach uses spatiotemporal characteristics from the route and previous bus trips to model short-term effects, and date/time variables and Gaussian processes for long-run forecasts. The model allows for flexible modeling of mean, variance and kurtosis spaces. We propose algorithms for Bayesian inference and for computing probabilistic forecast distributions. Experiments are performed using data from high-frequency buses in Stockholm, Sweden. 
Results show that Student-$t$ models outperform Gaussian ones in terms of log-posterior predictive power to forecast bus delays at specific stops, which reveals the importance of accounting for predictive uncertainty in model selection. Estimated Student-$t$ regressions capture typical temporal variability between within-day hours and different weekdays. Strong spatiotemporal effects are detected for incoming buses from immediately previous stops, which is in line with many  recently developed models. We finally show how Bayesian inference naturally allows for predictive uncertainty quantification, e.g. by returning the predictive probability that the delay of an incoming bus exceeds a given threshold. 

\end{abstract}


%
\IEEEpeerreviewmaketitle

\section{Introduction}\label{sec:intro}
%
%
%
%


\IEEEPARstart{A}{ccording} to United Nations demographic trends,  it is estimated that by 2030 cities will house $60\%$ of the world population \cite{un2018cities}, whereas the World Health Organization predicts $70\%$ by 2050 \cite{who2010urban}. This increased urbanization, along with the advent of automation and the Internet of Things, will consolidate the important role that urban public transport is currently playing in large urban areas worldwide, where the bulk of the population and economic activity is concentrated. Public transportation networks are a key element within the infrastructure of the city, not only by guaranteeing mobility and accessibility, but also by reducing private car usage and congestion, which has a positive impact on road safety and the environment, therefore contributing to a more sustainable urban development \cite{banister2008sustainable}. Also, public transport has an important social value as a facilitator of a universal, efficient, safe and sustainable mobility service for different social groups and communities \cite{un2020sust}. Despite the increasing transport demand and social benefits, the private car is still the preferred mode in many regions, and local authorities develop policies aimed at improving the image of public transportation and influencing the public's attitude towards its use \cite{beirao2007understanding}. Strategies to increase ridership can be taken from different angles. From the service perspective, keeping high standards in safety, availability and punctuality is essential. Advertisement plays also an important role in providing users accurate information about the public transportation system, and promoting its value and benefits.  Finally, administrative actions include e.g. the set up of congestion taxes, which revenues can be used for further investment in transport infrastructure \cite{eliasson2009stockholm}. 

Despite of all the above-mentioned strategies, even the most enthusiast users may change their minds if they becomes dissatisfied with the system. Attracting users to public transport is not enough, transport agencies should also aim to maximize retention. One aspect that determines the quality of public transport, and is therefore key to maximize user retention, is reliability. Transit reliability is a multifaceted concept that is often defined as the dependability of the service in terms of travel times (both waiting and riding), vehicle quality, safety, etc. \cite{ceder2016public}. Some authors have proposed measures to objectively quantify reliability from the passenger’s perspective, e.g. as the proportion of passengers receiving regular and punctual service \cite{barabino2015rethinking}\cite{barabino2016rethinking}. In a recent study \cite{van2018influences}, the authors review the different aspects that affect customer satisfaction and loyalty in public transportation. They find that, among multiple factors, users are particularly sensitive to unexpected waiting times, as it negatively affect their perception of the reliability of the system. This could be ameliorated by an effective and clear delivery of real-time information, which is especially important in urban, multi-modal networks where multiple-transfer journeys are common. Intelligent Transportation Systems (ITS) play a key role on how users perceive reliability through the use technologies like real-time information systems (RTI). One way to improve user's reliability perception of public transport with the existing technology, is by providing associated uncertainties in travel time predictions around the (mean) estimate. This approach of uncertainty quantification has recently gained attention in the public transport literature \cite{o2016uncertainty}\cite{yu2017using}, which comes as no surprise given the increasing need for accurate, and even personalized, forecasts within an highly stochastic environment. 

Accurate bus travel time predictions allow operators to perform real-time monitoring and take proactive control strategies, which eventually improve the service reliability by reducing passenger uncertainty at stops and vehicles \cite{ceder2016public}. Hence, in dealing with a crucial aspect of reliability, these models are relevant for both public transport users and operators. In this paper we develop a robust model for real-time bus travel time prediction that departs from Gaussian assumptions by using Student-$t$ errors. We define robustness as the model’s capacity to accommodate unusual or extreme observations in a data distribution \cite{gelman2013bayesian}. Our approach considers spatiotemporal characteristics from the route and previous bus trips, and allows for flexible modeling on mean, variance and degrees-of-freedom. We focus on schedule deviations at bus stops, although the proposed framework is flexible enough to accommodate models based on headway deviations, which are important in high-frequency services (see e.g. \cite{yu2016headway}\cite{yu2017probabilistic}\cite{barabino2019diagnosis}). We provide algorithms for Bayesian inference that enable the estimation of predictive uncertainty. Experiments are performed using data from buses in Stockholm, Sweden. 

The rest of the paper is organised as follows. Section \ref{sec:literature} reviews relevant literature on spatiotemporal bus travel time prediction and uncertainty estimation.  Section \ref{sec:methodology} presents the proposed model and inference methodology. The data and experimental setup are described in section \ref{sec:setup}, and results based on a route from Stockholm's bus network are discussed in section \ref{sec:results}. The last section presents the conclusions, limitations, and discuss some directions for future work. 
\section{Related Work}\label{sec:literature}
Bus travel time prediction is concerned with providing users and operators real-time information on arrivals, departures, and journey times for scheduled and running buses, given the current state of the network, the traffic conditions, and any other potentially useful data source. Whenever possible, a measure of uncertainty over the predictions is also desirable for the reasons stated above. Most ITS support travel time predictions mainly through GPS-enabled Automated Vehicle Location (AVL) technologies combined with e.g. road, traffic and passenger (e.g. smart cards) data. A large proportion of the literature is devoted to online models, since they provide real-time information for passengers and operators, although the prediction/estimation of bus arrival/running times can also be performed offline. Offline analysis using historical AVL records is useful to optimize bus route design and scheduling, see \cite{pili2019evaluating} for a recent review and comparative study. Models for travel time forecasting that are solely based on demand data are not common in the literature, although  \cite{zhou2017bus} successfully developed a regression model for predicting bus arrival times based on smart card swiping records and seat occupancy ratios. The spatiotemporal analysis and prediction of public transport demand is an active research topic, see e.g. \cite{mohamed2017clustering}\cite{qi2018analysis}\cite{karnberger2020network}, with potential applications to the travel time prediction problem. With respect to methodologies, the historical development is similar to that of road traffic forecasting \cite{vlahogianni2014short}, with more sophisticated, non-linear models from machine and deep learning becoming more popular as data availability and complexity increased. What follows is not an exhaustive survey on bus travel/arrival time prediction literature, but a brief review of works that are methodologically most relevant to our contribution, namely spatiotemporal and probabilistic.

Bus arrival time predictions at each route segment (i.e. the time interval between two consecutive stops) is performed in \cite{bin2006bus} using Support Vector Machines (SVM). The travel time is defined as a function of preceding bus times and segment information, whereas peak traffic conditions and weather are considered by training the model on different datasets. Though straightforward, their results illustrate the importance of considering spatial dependencies along the route. Also using SVM's, the work in \cite{yu2016headway} present stop-level headway (i.e. the time interval between two consecutive buses) prediction as a first step to bus-bunching detection. The model is fitted via least-squares to reduce the computational burden on large data, and uses previous headways, travel times and demand data as input.  Even though their proposed method does not yield probabilistic forecasts, stop-level information on headways is likely to reduce user's uncertainty, as pointed out by the authors. Deep learning has been recently used by \cite{pang2018learning} to tackle the bus arrival time prediction problem considering spatiotemporal factors. A long short-term memory (LSTM) neural network is fitted to large-scale bus data from Beijing to produce multi-step ahead forecasts. The structure of the recurrent neural network naturally captures the spatial dependencies and helps the model to outperform SVM and state-space models, among others, in mean prediction. The authors stress the importance of including meaningful route characteristics into the model, such as intersections and mid-points, rather than e.g. increasing spatial sampling. Another LSTM is trained in \cite{he2018travel} to model journeys which may include multiple bus trips. A single journey is divided in segments and transfer points, and predictions of riding and waiting times are done separately and then combined to produce the final forecast. As in the previous study, the LSTM network captures the network dependencies (e.g. between adjacent segments), and is able to outperform a number of off-the-shelf models. An interesting result is the variability of travel time uncertainty between routes, and how it increases for longer journeys due to transfers, intersections, etc. Recently, \cite{achar2019bus} proposed a spatial Kalman Filter (KF) to deal with the problem of arrival time prediction in a developing country, where the traffic conditions are very heterogeneous. The spatial and temporal correlations are explicitly learned first, and then embedded in a state space model for real-time prediction. Spatial dependencies are modeled in an autoregressive fashion based on travel times from previous segments, and time correlations as a non-linear function of recent bus travel times. An interesting aspect of their model is that correlations decay exponentially with time, which allows predictions to fall back to a steady-state if the available information is too distant.

Fewer studies, however, have explicitly addressed uncertainty quantification as a problem on its own. The authors in \cite{o2016uncertainty} show in a simple residual analysis how bus arrival times clearly depart from Gaussianity by exhibiting a heteroskedastic variance, skewness and even fat tails (high kurtosis). Under these circumstances, mean predictions are not appropriate, and quantile regression is proposed to quantify uncertainty. Gaussian processes (GP) and splines are both used successfully, although the performance of the former comes at a high computational cost. Survival models are used by \cite{yu2017using} to estimate bus travel times between bus stops and their associated uncertainties. The authors acknowledge the importance of uncertainty quantification as a mean to improve the perceived service reliability and set realistic expectations, as is not uncommon for real-time information systems to underestimate remaining times, as also pointed out by \cite{cats2016real}. In \cite{yu2017probabilistic} a Relevance Vector Machine is built to forecast bus headways based on demand data (smart cards). The Bayesian setting allows to provide probability intervals based on the posterior predictive distribution, although they recognize that the Gaussianity assumption is perhaps too restrictive. 

The main novelty of this work is the development of a robust regression model for probabilistic bus travel time prediction, and the necessary algorithms for its inference. To the best of our knowledge, this is the first study in transportation that develops such a model based on the Student-$t$ distribution with flexible regression modeling of the mean, variance and kurtosis. Five contributions are particularly noteworthy: i) our model explicitly departs from Gaussianity by assuming Student-$t$ errors; ii) flexibility via linear regressions on the model mean, scale and degrees of freedoms; iii) short-term and long-run forecasts via time-decaying spatiotemporal features and steady-state covariates; iv) efficient Metropolis-within-Gibbs sampler for Bayesian inference and generation of predictive densities; and v) probabilistic model comparison by evaluating the accuracy of the full predictive posterior distribution using the log-posterior predictive density (LPPD) instead of point-based measures.

\section{Methodology}\label{sec:methodology}
\subsection{Definitions}
Let a bus network be defined as a set of routes $R=\{r_1,\ldots,r_K\}$ operating simultaneously. We denote the $j$-th stop or station $s$ within bus route $k$ as $s_{kj}$, and define a route as an ordered set of stops $r_k = \{s_{k1},\ldots,s_{kJ_k}\}$, where $J_k$ is the number of stops in route $k$, and the direction of the journey is implicit in the ordering. The complete set of stops in the network would be $S=\cup_{k,j}s_{kj}$, with total number of stops $J=\sum_kJ_k$. We do not consider the physical distance between stops explicitly, but refer to that space interval as a segment.

Assume that we have access to a real-time flow of (actual) stop arrival times $T_{kj}$ from $K$ different bus routes, and the corresponding schedule times, $\tilde{T}_{kj}$. When an actual arrival time for stop $s_{kj}$ becomes available, the observed delay is
\begin{equation}\label{eq:delaydef}
    y_{kj}=T_{kj}-\tilde{T}_{kj}.
\end{equation}
If the incoming times are specified in seconds, (e.g. Unix time), the difference in \eqref{eq:delaydef} will just measure the deviation in seconds with respect to the original schedule at stop $s_{kj}$. Positive delay occurs when the bus is late, and negative when it arrives before the scheduled time. Since the flow of data is irregular across all variables, delay observations $y_{kj}(t)$ will only be available on intervals $t$ where there was an actual arrival time for that route and stop and, as expected, the sparsity of the data increases with the time resolution. We have not considered the unseen overtaking of buses when creating the delays, as bus overtakes are rare in our data. Data irregularity has been tackled in a model-based fashion, as detailed in the next section. Note that inconsistencies in AVL records (e.g. missing data, overtakes) may have an important effect on regularity (i.e. headway-based) measures, in contrast to punctuality measures such as bus delay at stops, see e.g. \cite{barabino2017time}. We will work with discrete time at the minute level so that every delay will be assigned to a specific within-day minute, $t\in\{1,\ldots,1440\}$, depending on the minute the actual arrival time occurred. 
\begin{table}
\caption{Notation}
\begin{tabular}{|c|c|}
\hline
 $k=1,\ldots,K$ & Route index\\
 $j=1,\ldots,J$ & Stop index\\
 $i=1,\ldots,N$ & Data observation index\\
 $t\in\{1,\ldots,1440\}$ & Time index (minute within a day)\\
 $r_{k}$ & $k$-th route\\
 $K$ & Total number of routes\\
  $R\in \{r_1,\ldots,r_K\}$ & Complete set of routes, $R=\cup_{k}r_{k}$\\
 $s_{kj}$ & $j$-th stop at route $k$\\
 $J_k$& Number of stops at route $k$ \\
 $J$& Total number of stops, i.e. $\sum_kJ_k$ \\
 $S\in \{s_1,\ldots,s_J\}$ & Complete set of stops, $S=\cup_{k,j}s_{kj}$\\
  $N_j$& Number of observations at stop $j$ \\
 $N$& Total number of observations, i.e. $\sum_jN_j$\\
 $y_{kj}(t)$& Delay (sec.) at stop $j$ from route $k$ at time $t$\\ 
 $f(\cdot)$ & Long-run (steady-state) delay function\\
 $g(\cdot)$ & Short-run delay function\\
 $h_j(t,l)$ & Time of $j$-th most recent delay of bus $l$ \\
 $\delta_j$ & Geometric time-decaying factor at stop $j$\\
 $P$& Number of previous data points in $g$ function\\
 $L$& Number of preceding buses in $g$ function\\
 $\mathcal{G}^*_j$ & Set of $j$'s preceding $P$ stops from graph $\mathcal{G}$\\
 $\boldsymbol{Z,W}$ & Steady-state and short-run feature matrices\\
 $\boldsymbol{X}$ & Combined, $\boldsymbol{X}=(\boldsymbol{Z},\boldsymbol{W})$, feature matrix\\
 $\boldsymbol{\alpha,\gamma}$ & Steady-state and short-run parameter vectors\\
$\boldsymbol{\beta}$ & Combined, $\boldsymbol{\beta}=(\boldsymbol{\alpha}^\top,\boldsymbol{\gamma}^\top)^\top$, model parameters\\
 $\boldsymbol{\theta}$ & Generic parameter vector\\
 $\mathcal{N}(\mu,\sigma^2)$ & Normal distribution with mean $\mu$ and variance $\sigma^2$\\
 $\mathcal{T}(\mu,\sigma^2,\nu)$ & Student-$t$ distribution with location $\mu$, scale $\sigma^2$\\
 &  and degrees of freedom $\nu$\\
 $\mathcal{N}(\boldsymbol{\mu},\Sigma)$ & Multivariate Normal distribution with mean $\boldsymbol{\mu}$\\
 & and covariance matrix $\Sigma$\\
 $\mathcal{T}(\boldsymbol{\mu},\Sigma,\nu)$ & Multivariate Student-$t$ distribution with location $\boldsymbol{\mu}$,\\
 & covariance matrix $\Sigma$ and degrees of freedom $\nu$\\
  $\text{Inv-}\chi^2(\nu,\tau^2)$ & Scaled inverse-chi-square distribution with\\
  & degrees of freedom $\nu$ and scale $\tau^2$\\
 $\odot, \oslash$ & Hadamard (elementwise) product and division\\
 $\psi(\cdot), \psi_1(\cdot)$ & Digamma and trigamma functions\\
 $\mathbf{g}(\cdot), \mathbf{H}(\cdot)$ & Gradient and Hessian functions\\
  \hline
\end{tabular}
\end{table}

\begin{figure}
    \centering
    \includegraphics[width=3.5in]{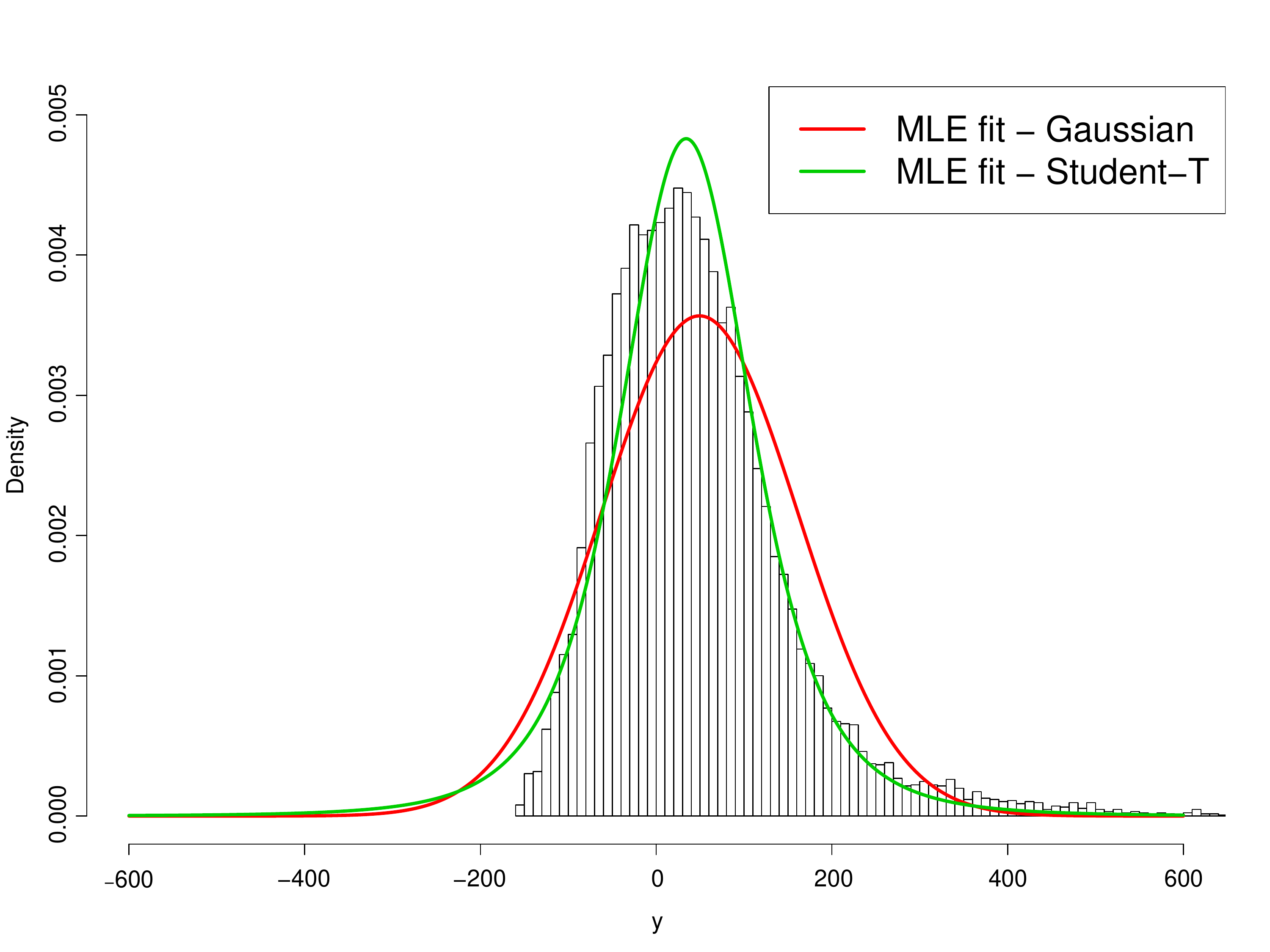}
    \includegraphics[width=3.5in]{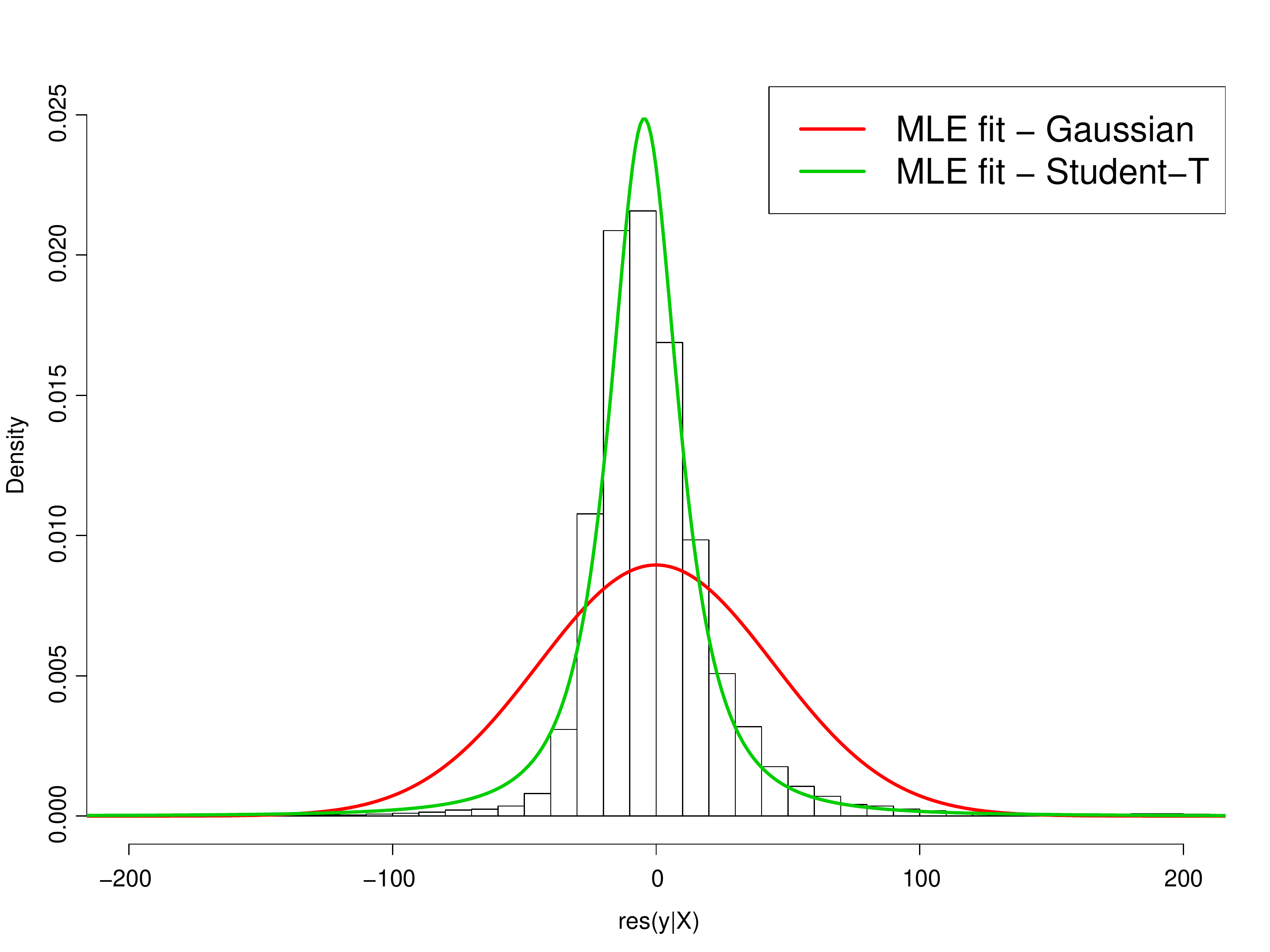}
    \caption{Delay distribution for a random stop (stop 8 at route 89 - top), and the residual distribution from a linear model for delays at the same stop based on delays from previous stops in the route (bottom). Both figures suggest a heavy-tailed model such as e.g. the Student-$t$ distribution.}
    \label{fig:hist_delay_residuals}
\end{figure}
\subsection{General model}\label{sec:general_model}
A commonly used linear model for multivariate time series is the vector autoregressive (VAR) model of order $L$ \cite{lutkepohl2005new}
\begin{equation}\label{eq:VAR}
    \boldsymbol{y}_t=\boldsymbol{\mu}+\Pi_1\boldsymbol{y}_{t-1}+\ldots+\Pi_L\boldsymbol{y}_{t-L} + \boldsymbol{\varepsilon}_t,
\end{equation}
where $\boldsymbol{\varepsilon}_t\sim\mathcal{N}(0,\boldsymbol{\Sigma})$ is an independent sequence of vector disturbances and $\{\boldsymbol{\mu},\Pi_1,\ldots,\Pi_L\}$ the VAR parameters. There are however several complications with using a standard VAR model for modeling the vector of bus delays at all stops, $\boldsymbol{y}_t$. 

First, VAR models are heavily overparameterized for all but the smallest of bus networks. The large number of parameters can be reduced using spatial constraints by representing the set of consecutive stops in a route as vertices in a directed graph $\mathcal{G}^k$ and keeping only the elements in the matrices $\Pi_1\ldots\Pi_L$ that correspond to actual edges within the graph. Second, bus delay data tend to be heteroskedastic and heavy-tailed, so the homoskedastic Gaussian disturbance need to be generalized to allow for a non-constant variance and kurtosis over the day and possibly also depending on recently observed delays. Third, and most difficult to solve in a VAR context, is that the event-based nature of bus records gives a very large proportion of missing observations in $\boldsymbol{y}_t$. This is problematic in VARs as we would have also to resort to imputation to fill in the missing bus information in order to perform inference, which is unrealistic. Finally, the model should use recently observed data for short-run prediction, and also converge to some well-defined steady-state distribution at longer forecast horizons to make it usable for route planning. The steady-state VAR in \cite{villani2009steady} achieves this effect for the mean, but not for the variance and kurtosis. We therefore propose to instead use a regression model on the stop level with spatial and temporal features using discounting of past data similar to the approach in \cite{achar2019bus}. Using a regression model with time discounted features also makes it easy to model the long-run predictive distribution and opens up for flexible modeling of heteroskedasticity and heavy-tailed disturbances.

The proposed model applies to any set of stops $S$ regardless of its route, so we drop the route index $k$ for clarity in the exposition. The conditional distribution of the delay at time $t$ for a bus arriving to stop $j$, $y_j(t)$, given a feature vector $\mathbf{x}_j(t)$, is modeled as
\begin{equation}\label{eq:model_general}
    y_j(t)|\mathbf{x}_j(t)\sim \mathcal{T}\big[\mu(\mathbf{x}_j(t)),\sigma^2(\mathbf{x}_j(t)), \nu(\mathbf{x}_j(t)) \big],
\end{equation}
where $\mathcal{T}(\mu,\sigma^2,\nu)$ is the Student-$t$ distribution with degrees of freedom $\nu>0$, location $\mu$ and scale $\sigma^2>0$ \cite{gelman2013bayesian}. The Student-$t$ has density function
\begin{equation}
    p(y|\mu,\sigma^2,\nu)=\frac{\Gamma((\nu+1)/2)}{\Gamma(\nu/2)\sigma\sqrt{\nu\pi}}\left(1+\frac{(y-\mu)^2}{\nu\sigma^2}\right)^{-(\nu+1)/2},
\end{equation}
where $\Gamma(\cdot)$ is the gamma function. If $Y\sim \mathcal{T}(\mu,\sigma^2,\nu)$ then $E[Y]=\mu$ and $V[Y]=\nu\sigma^2/(\nu-2)$ for $\nu>1$ and $\nu>2$, respectively, and $\nu$ controls the tail thickness such that the distribution converges to $Y\xrightarrow{d}\mathcal{N}(\mu,\sigma^2)$ as $\nu\rightarrow\infty$. Figure \ref{fig:hist_delay_residuals} shows that a Student-$t$ distribution seems to capture bus delay residuals well.

\subsection{Feature construction}
The Student-$t$ model parameters $\mu$, $\sigma^2$ and $\nu$ in \eqref{eq:model_general} are all functions of the time indexed feature vector $\mathbf{x}_j(t)$. As detailed below, $\mathbf{x}_j(t)$ contains functions of time and time-discounted past delays in the network to capture the dynamics in time and space. Different features may be used in $\mu$, $\sigma^2$ and $\nu$. 

For each of $\mu$, $\sigma^2$ and $\nu$ we use a short-term and a long-term component. For example, the mean in \eqref{eq:model_general} is modeled by
\begin{equation}
    \mu(\mathbf{x}_j(t)) = f_j(t)+g_j(t)
\end{equation}
with long-run part
\begin{equation}\label{eq:longterm}
    f_j(t) = \mathbf{z}_j(t)^\top\boldsymbol{\alpha}_j
\end{equation}
and short-run effects modeled by $g_j(t)$. Importantly, the short-run effects will be modeled so that $g_j(t)\rightarrow0$ when forecasting at longer horizons. The term $f_j(t)$ will therefore be referred to as the steady-state and will be the determinant of forecasts at longer horizons, e.g. predictions much later in the day or even several days ahead for route planning. The steady-state feature vector $\mathbf{z}_j(t)$ models additive effects for the within-day congestion peaks (hourly), within-week differences (day of the week) and seasonal patterns. We model some of these effects non-parametrically using Gaussian process priors \cite{rasmussen2006gaussian} on the parameters $\boldsymbol{\alpha}_j$ to encode smoothness over e.g. the hours of the day. 

\begin{figure}[ht]
\centering
\includegraphics[height=3.2in]{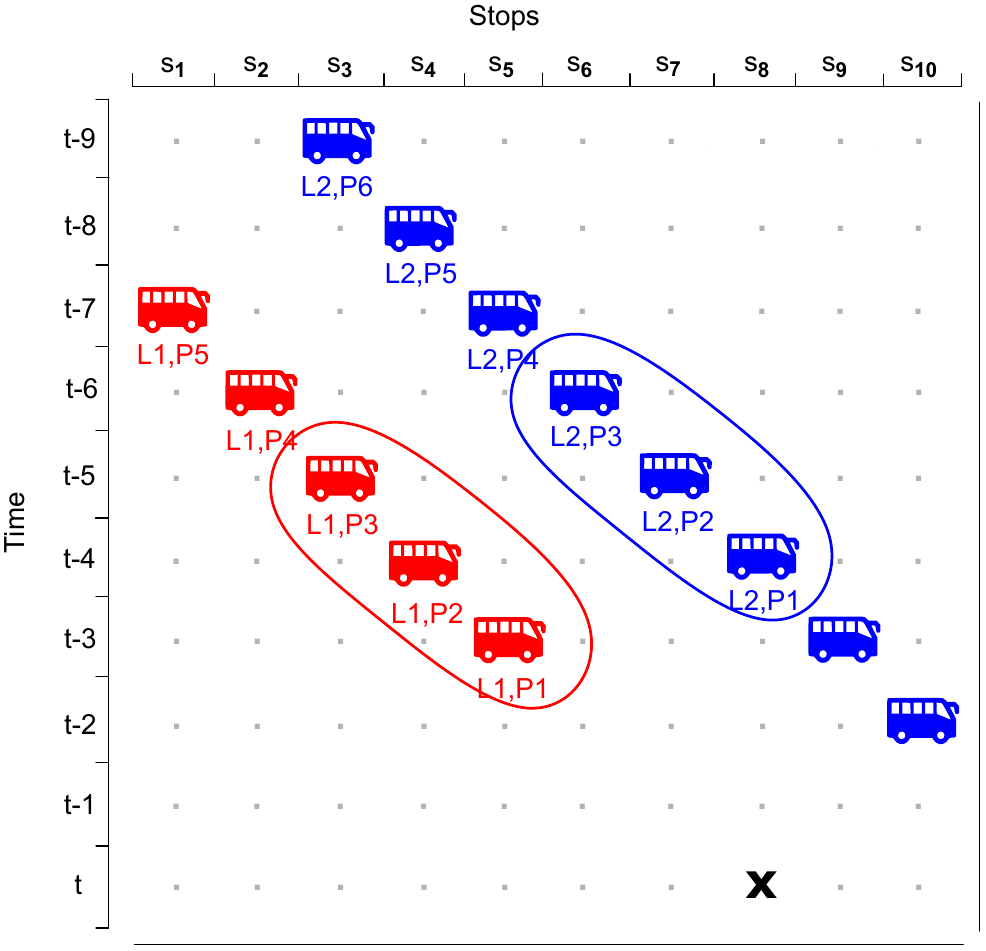}
\caption{Spatiotemporal traces of two buses traversing a hypothetical route with ten stops. We want to predict the next delay at stop $j=8$ and time $t$ (black cross), using $P=3$ previous delays from the $L=2$ most recent buses. The observations used to construct the model features are encircled. The sets of preceding stops would be $\mathcal{G}^*_j(t,l=1)=\{s_5,s_4,s_3\}$ and $\mathcal{G}^*_j(t,l=2)=\{s_8,s_7,s_6\}$, for the incoming and previous buses, respectively. }
\label{fig_busfeatures}
\end{figure}

In a similar spirit as in \cite{dessouky1999bus}, we model the short-term part of the mean, $g_j(t)$, as a linear combination of $P$ previous delays from $L$ recent buses coming to stop $j$. The set of recent buses include the incoming and $L-1$ previous buses. The $P$ preceding stops in the route graph from which delay data is available are $\mathcal{G}^*_j(t,l)=\{{s_{j-d}},s_{j-d-1},\ldots,s_{j-d-(P-1)}\}$, $0\le d\le j-P$, for every bus $l$, see Figure \ref{fig_busfeatures} for an illustration. Following \cite{achar2019bus} we use time-discounting to make sure that the effect of delays observed in the distant past are shrunk toward zero. The discounting also gives us the desired effect of reverting to the steady-state $f_j(t)$ when forecasting at longer horizons. Specifically, we model the short-run effects by 
\begin{equation}\label{eq:mean_model}
g_j(t)=\sum\limits_{l=1}^{L}\sum\limits_{p\in\mathcal{G}^*_j(t,l)}y_p(h_p(t,l))\delta_j^{t-h_p(t,l)}\gamma_{jpl},
\end{equation}
where the function $h_j(t,l)$ returns the time of the $j$-th previous observation of bus $l$ with respect to the current time $t$, while the term involving $\delta\in [0,1]$ geometrically shrinks observed past delays as they move away from $t$. The regression coefficients $\gamma_{jpl}$ capture the strength of the effect of the $p$-th last observation from the $l$-th previous bus in predicting a delay at stop $j$ at any time $t$.

Note that in the hypothetical scenario of a constant flow of data, i.e. having delay observations $y_j(t)$ for every $j$ and $t$ (e.g. every minute), we have $h_p(t,l)=t+p+1-l-j=\kappa$ for all $p\in\mathcal{G}^*_j(t,l)$ and the short-term part of the proposed model reduces to a VAR($L$) with spatial constrains from $P$ preceding stops in the route graph $\mathcal{G}^*_j=\{{s_j},s_{j-1},\ldots,s_{j-P}\}$
\begin{equation}
g_j(t)=\sum\limits_{l=1}^{L}\sum\limits_{p\in\mathcal{G}^*_j}y_p(\kappa)\tilde{\gamma}_{jpl},
\end{equation}
where $\tilde{\gamma}_{jpl}=\delta_j^{t-\kappa}\gamma_{jpl}$ are the VAR parameters.

Now given a data sample of size $N=\sum_j{N_j}$ containing delays for $J$ stops $y=\{y_j(t_1),\cdots, y_j(t_{N_j})\}_{j=1}^J$, we can write the mean $\boldsymbol{\mu}_j = E[\boldsymbol{y}_j]$  for a selected stop $j$ as
\begin{equation}\label{eq:obs_model}
\boldsymbol{\mu}_j =\mathbf{Z}^{\mu}_j\boldsymbol{\alpha}^{\mu}_j+\textbf{W}^{\mu}_j\boldsymbol{\gamma}^{\mu}_j,
\end{equation}
where a typical row of the steady-state feature matrix $\mathbf{Z}^{\mu}_j$ is $\mathbf{z}_j(t)$ from \eqref{eq:longterm} and a typical element of the short-run features in $\textbf{W}^{\mu}_j$ is a time discounted past delay in stop $j$ or one of its neighboring stops: $y_p(h_p(t,l))\delta_j^{t-h_p(t,l)}$ for some $l=1,...,L$ and $p\in \mathcal{G}_j^*(t,l)$.

\begin{figure}[hb]
	\centering
	\includegraphics[width=3.5in]{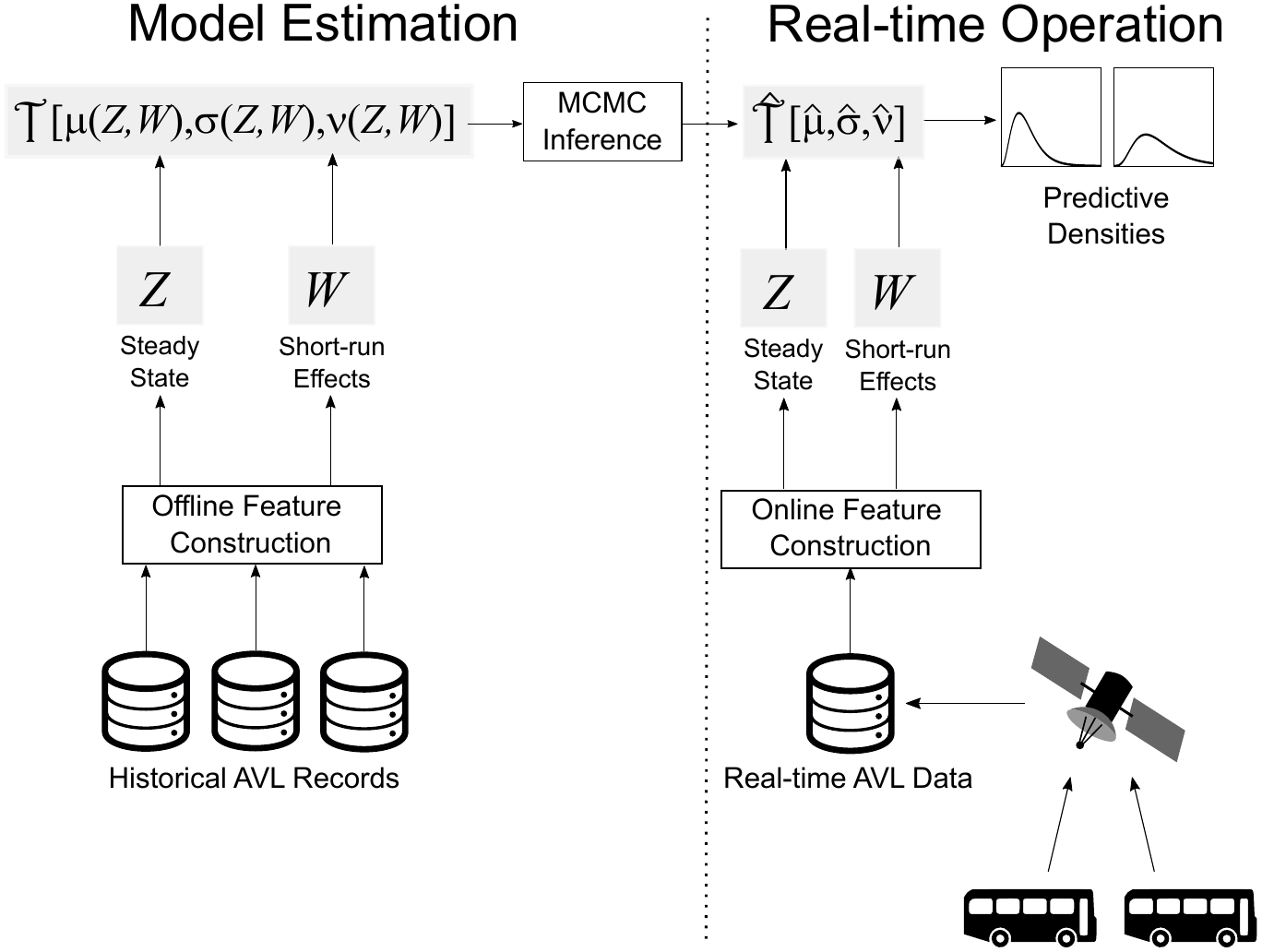}
	\caption{Illustrative overview of the methodological framework.}
	\label{fig:flowchart_model}
\end{figure}

We account for heteroskedasticity by modeling the logarithm of the scale and degrees of freedom as linear regressions in a similar fashion as for the model mean
\begin{eqnarray}
    \ln\boldsymbol{\sigma}^2_j=\mathbf{Z}^{\sigma}_j\boldsymbol{\alpha}^{\sigma}_j+\textbf{W}^{\sigma}_j\boldsymbol{\gamma}^{\sigma}_j\label{eq:noise_model}\\
    \ln\boldsymbol{\nu}_j=\mathbf{Z}^{\nu}_j\boldsymbol{\alpha}^{\nu}_j+\textbf{W}^{\nu}_j\boldsymbol{\gamma}^{\nu}_j,\label{eq:df_model}
\end{eqnarray}
where $\mathbf{Z}$ and $\mathbf{W}$ are again the steady-state and short-term features, respectively. We use the same features for the steady-state as used in the mean function. For the short-term variations in $\boldsymbol{\sigma}^2_j$ and $\boldsymbol{\nu}^2_j$ we use the absolute differences of previous delays as features instead of the delays used in the mean. Figure \ref{fig:flowchart_model} presents an overall illustration of the methodology. Note that the general model takes all features for the regressions on mean, scale, and degrees-of-freedom. However, it is possible to work with benchmark models that are simpler variants of the general model by removing/simplifying some components, as we explain in section \ref{sec:benchmark_models}. 

\subsection{Likelihood and prior distributions}
In this section we denote the observed delays at a given bus stop simply as $\boldsymbol{y}$, hence dropping route, stop and time indexes for clarity. The model for a given stop can be written as
\begin{equation}
    \boldsymbol{y} = \boldsymbol{\mu} + \boldsymbol{\sigma}^2\odot\boldsymbol{T},
\end{equation}
where $\boldsymbol{T}\sim \mathcal{T}(0,1,\boldsymbol{\nu})$, the column vectors $\boldsymbol{y},\boldsymbol{\mu},\boldsymbol{\sigma}^2$, and $\boldsymbol{\nu}$ are of size $n$, and $\odot$ is the Hadamard (elementwise) product. As detailed above, the mean, scale and degrees of freedom are dependent on covariate information
\begin{eqnarray}
    \boldsymbol{\mu}=\mathbf{X}_{\mu}\boldsymbol\beta_{\mu}\\
    \ln\boldsymbol{\sigma}^2=\mathbf{X}_{\sigma}\boldsymbol\beta_{\sigma}\\
    \ln\boldsymbol{\nu}=\mathbf{X}_{\nu}\boldsymbol\beta_{\nu},
\end{eqnarray}
where $\mathbf{X}_{\mu} = (\mathbf{Z}_{\mu},\mathbf{W}_{\mu})$ and $\boldsymbol\beta_{\mu}=(\boldsymbol\alpha_{\mu}^\top,\boldsymbol\gamma_{\mu}^\top)^\top$, and similarly for $\boldsymbol{\sigma}^2$ and $\boldsymbol{\nu}$.

To facilitate efficient posterior computations, we use the mixture-of-normal's representation of the Student-$t$ distribution \cite{gelman2013bayesian}, which for each data point in the sample, $i=1,\ldots,n$, becomes
\begin{eqnarray}
    y_i \sim \mathcal{N}(\mu_i,\alpha^2U_i)\label{eq:scale_mixture1}\\
    U_i \sim \text{Inv-}\chi^2(\nu_i,\tau_i^2).\label{eq:U_prior}\label{eq:scale_mixture2}
\end{eqnarray}
Note that the pseudo-variances $\boldsymbol{U}$, and scale parameters $\boldsymbol{\tau}$ and $\alpha^2$ are not identified in the specification above. However, $\boldsymbol{\mu}$, $V_i=\alpha^2U_i$, $\boldsymbol{\sigma}^2=\alpha^2\boldsymbol{\tau}^2$, and $\boldsymbol{\beta}_{\mu},\boldsymbol{\beta}_{\sigma},\boldsymbol{\beta}_{\nu}$ are identified. The purpose of the scalar parameter $\alpha$ is to prevent the sampler to get stuck in low values of $\boldsymbol{U},\boldsymbol{\tau}$, and has no meaning on its own. If we denote the vector of all model parameters as $\boldsymbol{\theta}$, the complete-data likelihood is
\setlength{\arraycolsep}{0.1em}
\begin{eqnarray}
    p(\boldsymbol{y}|\boldsymbol{\theta})&=&\prod_{i=1}^{n}\mathcal{N}(y_i;\mu_i,\alpha^2U_i)\text{Inv-}\chi^2(U_i;\nu_i,\tau_i^2).
\end{eqnarray}
The prior distribution for the augmented data $\boldsymbol{U}$ is given by \eqref{eq:U_prior}, and for $\alpha$ we use a non-informative uniform prior on log scale. The prior distributions for the mean, scale and degrees of freedom regression parameters are
\begin{IEEEeqnarray}{l}
  \boldsymbol{\beta}_{\mu}\sim \mathcal{N}(\boldsymbol{m}_{\mu},S_{\mu})\\
  \boldsymbol{\beta}_{\sigma}\sim \mathcal{N}(\boldsymbol{m}_{\sigma},S_{\sigma})\\
  \boldsymbol{\beta}_{\nu}\sim \mathcal{N}(\boldsymbol{m}_{\nu},S_{\nu}).
\end{IEEEeqnarray}
Unless otherwise stated, we will place non-informative flat priors on the different $\boldsymbol{\beta}$'s, but enforce smoothness in subsets of parameters corresponding to time-dependent covariates such as the hour of the day. In that case we use a Gaussian process prior, i.e. $\boldsymbol{\beta}\sim\mathcal{GP}(0,\sigma(\cdot,\cdot))$, with a squared-exponential covariance kernel $\sigma(x,x')=\exp(-0.5(x-x')^2/l^{2})$ with predefined lengthscale $l$. The prior covariance matrices $S$ will therefore contain a block with the Gram matrix that encodes the prior dependencies between the corresponding parameters.  


\subsection{Metropolis-within-Gibbs sampler}\label{sec:gibbs_sampler}
\subsubsection{Full conditional posterior for $U_i$}
The conditional posterior distribution for the pseudo-variances follow conjugate results, and therefore each $U_i$ can be drawn independently from 
\begin{equation}
    U_i|- \sim \text{Inv-}\chi^2\left(\nu_i+1,\frac{\nu_i\tau_i^2+((y_i-x_{\mu i}\beta_{\mu})/\alpha)^2}{\nu_i+1}\right),
\end{equation}
where $\vert-$ is short-hand notation for conditioning on all other parameters and the data.    
\subsubsection{Full conditional posterior for $\boldsymbol{\beta}_{\mu}$}
Conditioned on the rest of the parameters, the normal prior for $\boldsymbol{\beta}_{\mu}$ is conjugate. Define $\tilde{X}=\boldsymbol{\omega}\odot X_{\mu}$ and $\tilde{\boldsymbol{y}}=\boldsymbol{\omega}\odot \boldsymbol{y}$, where $\boldsymbol{\omega}=(\alpha^2\boldsymbol{U})^{-1/2}$. The regression parameters for the mean $\boldsymbol{\mu}$ can then be drawn from
\begin{IEEEeqnarray}{l}
  \boldsymbol{\beta}_{\mu}|-\sim \mathcal{N}(\boldsymbol{\mu}_{\mu}^*,\Sigma_{\mu}^*)\\
(\Sigma_{\mu}^*)^{-1}=S_{\mu}^{-1}+\tilde{X}^T\tilde{X}\\
\boldsymbol{\mu}_{\mu}^*=\Sigma_{\mu}^*(\tilde{X}^T\tilde{\boldsymbol{y}}+S_{\mu}^{-1}\boldsymbol{m}_{\mu}).
\end{IEEEeqnarray}
\subsubsection{Full conditional posterior for $\boldsymbol{\beta}_{\sigma}$}
The conditional posterior is proportional to
\begin{IEEEeqnarray}{LL}
    \boldsymbol{\beta}_{\sigma}|-\propto&\mathcal{N}(\boldsymbol{\beta}_{\sigma};\boldsymbol{m}_{\sigma},S_{\sigma})\times\nonumber\\
    &\prod_{i=1}^{n}\text{Inv-}\chi^2(U_i;\nu_i,\exp(\boldsymbol{x}^T_{\sigma i}\boldsymbol{\beta}_{\sigma})),
\end{IEEEeqnarray}
which is not of any known form. We follow \cite{li2010flexible,villani2009regression} and perform a Metropolis-Hastings step using finite-step Newton proposals to sample from the conditional posterior of $\boldsymbol{\beta}_{\sigma}$. Given a current sample of $\boldsymbol{\beta}_{\sigma}$ we can iterate a few steps in the Newton's algorithm, i.e.
\begin{equation}
    \boldsymbol{\beta}^{k+1}=\boldsymbol{\beta}^{k}- \mathbf{H}(\boldsymbol{\beta}^{k})^{-1}\mathbf{g}(\boldsymbol{\beta}^{k}),
\end{equation}
in order to approximate the posterior mode and covariance matrix, provided the gradient $\mathbf{g}$ and Hessian $\mathbf{H}$ are available in closed form. Expressions for both $\mathbf{g}$ and $\mathbf{H}$ are derived below. Once we have an estimate of the mode, $\hat{\boldsymbol{\beta}}$, the proposal is sampled from
\begin{IEEEeqnarray}{ll}
    \boldsymbol{\beta}_{\sigma}^{\text{new}}|\boldsymbol{\beta}_{\sigma}^{\text{old}}\sim \mathcal{T}\left(\hat{\boldsymbol{\beta}},-\mathbf{H}(\hat{\boldsymbol{\beta}})^{-1},d\right),
\end{IEEEeqnarray}
where the proposal distribution $q\sim\mathcal{T}(\mu,\Sigma,d)$ is the multivariate Student-$t$ with $d$ degrees of freedom, location $\mu$ and covariance matrix $\Sigma$. In practice, setting $d=10$, and taking just 2 or 3 Newton iterations is enough for an efficient sampling. The acceptance probability ratio is given by
\begin{IEEEeqnarray}{ll}
    p_{\text{accept}}=\min\left(1,\frac{p(\boldsymbol{y}|\boldsymbol{\beta}_{\sigma}^{\text{new}})q(\boldsymbol{\beta}_{\sigma}^{\text{old}}|\boldsymbol{\beta}_{\sigma}^{\text{new}})}{p(\boldsymbol{y}|\boldsymbol{\beta}_{\sigma}^{\text{old}})q(\boldsymbol{\beta}_{\sigma}^{\text{new}}|\boldsymbol{\beta}_{\sigma}^{\text{old}})}\right).
\end{IEEEeqnarray}

The gradient and Hessian for the log-posterior of $\boldsymbol{\beta}_{\sigma}$ are obtained by repeated use of the chain rule, yielding
\begin{IEEEeqnarray}{lLl}
    \mathbf{g}(\boldsymbol{\beta}_{\sigma})&=&\frac{\partial \ln p(\boldsymbol{\beta}_{\sigma}|-)}{\partial \boldsymbol{\beta}_{\sigma}}\nonumber\\
    &=&\frac{1}{2}[X_{\sigma}^T\boldsymbol{\nu}-X_{\sigma}^T(\boldsymbol{\nu}\odot\exp(X_{\sigma}\boldsymbol{\beta}_{\sigma})\oslash\boldsymbol{U})]\nonumber\\
    &-&S_{\sigma}^{-1}(\boldsymbol{\beta}_{\sigma}-\boldsymbol{m}_{\sigma})
\end{IEEEeqnarray}
\begin{IEEEeqnarray}{lLl}
    \mathbf{H}(\boldsymbol{\beta}_{\sigma})&=&\frac{\partial^2 \ln p(\boldsymbol{\beta}_{\sigma}|-)}{\partial \boldsymbol{\beta}_{\sigma}\partial\boldsymbol{\beta}_{\sigma}^T}\nonumber\\
    &=&-\frac{1}{2}X_{\sigma}^T(\boldsymbol{\nu}\odot\exp(X_{\sigma}\boldsymbol{\beta}_{\sigma})\oslash \boldsymbol{U})X_{\sigma}-S_{\sigma}^{-1}.
\end{IEEEeqnarray}
\subsubsection{Full conditional posterior for $\boldsymbol{\beta}_{\nu}$}
Analogously as for $\boldsymbol{\beta}_{\sigma}$, the conditional posterior 
\begin{IEEEeqnarray}{LL}
    \boldsymbol{\beta}_{\nu}|-\propto&\mathcal{N}(\boldsymbol{\beta}_{\nu};\boldsymbol{m}_{\nu},S_{\nu})\times\nonumber\\
    &\prod_{i=1}^{n}\text{Inv-}\chi^2(U_i;\exp(x_{\nu i}\boldsymbol{\beta}_{\nu}),\tau_i^2)
\end{IEEEeqnarray}
is not of any known form. We follow the same strategy described above for $\boldsymbol{\beta}_{\sigma}$, using the following gradient and Hessian for the Newton-based Metropolis-Hastings sampling step of the regression coefficients $\boldsymbol{\beta}_{\nu}$,
\begin{IEEEeqnarray}{lLl}
    \mathbf{g}(\boldsymbol{\beta}_{\nu})&=&\frac{\partial \ln p(\boldsymbol{\beta}_{\nu}|-)}{\partial\boldsymbol{\beta}_{\nu}}\nonumber\\
    &=&\frac{1}{2}[X_{\nu}^T(\hat{\boldsymbol{\nu}}\odot\ln(\hat{\boldsymbol{\nu}}/2)+\hat{\boldsymbol{\nu}})\nonumber\\
    &-&X_{\nu}^T(\hat{\boldsymbol{\nu}}\odot\psi(\hat{\boldsymbol{\nu}}/2))+X_{\nu}^T(\ln\boldsymbol{\tau}^2\odot\hat{\boldsymbol{\nu}})\nonumber\\
    &-&X_{\nu}^T(\ln \boldsymbol{U}\odot\hat{\boldsymbol{\nu}})-X_{\nu}^T(\hat{\boldsymbol{\nu}}\odot\boldsymbol{\tau}^2\oslash\boldsymbol{U})]\nonumber\\
    &-&S_{\nu}^{-1}(\boldsymbol{\beta}_{\nu}-\boldsymbol{m}_{\nu})
\end{IEEEeqnarray}
\begin{IEEEeqnarray}{lLl}
    \mathbf{H}(\boldsymbol{\beta}_{\nu})&=&\frac{\partial^2 \ln p(\boldsymbol{\beta}_{\nu}|-)}{\partial\boldsymbol{\beta}_{\nu}\partial\boldsymbol{\beta}_{\nu}^T}\nonumber\\
    &=&\frac{1}{2}X_{\nu}^T[\text{diag}(2\hat{\boldsymbol{\nu}}+\hat{\boldsymbol{\nu}}\odot\ln(\hat{\boldsymbol{\nu}}/2))\nonumber\\
    &-&\text{diag}(\hat{\boldsymbol{\nu}}\odot(\psi(\hat{\boldsymbol{\nu}}/2)+\hat{\boldsymbol{\nu}}/2\odot\psi_1(\hat{\boldsymbol{\nu}}/2)))\nonumber\\
    &+&\text{diag}(\hat{\boldsymbol{\nu}}\odot\ln\boldsymbol{\tau}^2)-\text{diag}(\hat{\boldsymbol{\nu}}\odot\ln\boldsymbol{U})\nonumber\\
    &-&\text{diag}(\hat{\boldsymbol{\nu}}\odot\boldsymbol{\tau}^2\oslash\boldsymbol{U})]X_{\nu}-S_{\nu}^{-1},
\end{IEEEeqnarray}
where $\hat{\boldsymbol{\nu}}=\exp(X_{\nu}\boldsymbol{\beta}_{\nu})$, and $\psi(\cdot),\psi_1(\cdot) $ are the digamma and trigamma functions, respectively, and $\oslash$ the Hadamard (elementwise) division.

\subsubsection{Full conditional posterior for $\alpha^2$}
Conditional on everything else, $\alpha^2$ is just a normal variance parameter with full conditional posterior as
\begin{equation}
    \alpha^2|- \sim \text{Inv-}\chi^2\left(n,\frac{1}{n}\sum_{i=1}^n\frac{(y_i-x_{\mu i}\beta_{\mu})^2}{U_i}\right).
\end{equation}

\subsection{Benchmark models}\label{sec:benchmark_models}
Seven models will be used in our experiments, featuring different noise assumptions and structure. 
All benchmark models, with the exception of the random-walk, are variants of the general model presented in \ref{sec:general_model}. Table \ref{tab:modelinfo} presents a summary of the different covariates entering in each model, and a detailed description follows. For simplicity, we denote the mean covariate information and parameters for a model at stop $j$ as $\mathbf{X}^{\mu}_j = (\mathbf{Z}^{\mu}_j,\mathbf{W}^{\mu}_j)$ and $\boldsymbol\beta^{\mu}_j=(\boldsymbol\alpha^{\mu\top}_j,\boldsymbol\gamma^{\mu\top}_j)^\top$, and similarly for $\boldsymbol{\sigma}^2$ and $\boldsymbol{\nu}$.  
\subsubsection{Historical average}
We define our historical average as the steady state for the model mean $\boldsymbol\mu_j$, at a given stop $j$, i.e. 
\begin{IEEEeqnarray}{lLl}\label{eq:model_ha}
    \boldsymbol{y}_j|\mathbf{Z}^{\mu}_j\sim \mathcal{N}(\mathbf{Z}^{\mu}_j\boldsymbol\gamma^{\mu}_j,\sigma^2_j),
\end{IEEEeqnarray}
where the noise is Gaussian and homoskedastic. Posterior sampling is straightforward by using conjugate results for linear models, e.g. normal-inverse-gamma prior for  $\boldsymbol\gamma^{\mu}_j$ and $\sigma^2_j$.
\subsubsection{Random walk}
A random walk will be centered on the most recent delay available for the incoming bus, which is $h$ seconds distant from the current time $t$. We assume a Gaussian noise with a variance that increases linearly with the distance $h$,
\begin{IEEEeqnarray}{lLl}\label{eq:model_rw}
 {y}_j(t)| {y}_j(t-h)\sim\mathcal{N}({y}_j(t-h),h\sigma^2_j).
\end{IEEEeqnarray}
The noise variance is estimated from the transformed data $\tilde{y}=({y}_j(t)- {y}_j(t-h))/\sqrt{h}\sim\mathcal{N}(0,\sigma^2_j)$. 
\subsubsection{Gaussian homoskedastic}
The first model that uses all effects is Gaussian, centered at a mean with the steady-state and short-term effects defined in Eqs \eqref{eq:longterm}-\eqref{eq:mean_model}, and has a homoskedastic variance. We sample from the conditionally conjugate posteriors of $\boldsymbol\beta^{\mu}_j,\sigma^2_j$ using Gibbs sampling. 
\begin{IEEEeqnarray}{lLl}\label{eq:model_gauss_homo}
    \boldsymbol{y}_j|\mathbf{X}^{\mu}_j\sim \nonumber \mathcal{N}(\boldsymbol{\mu}_j,\sigma^2_j)\\
\boldsymbol{\mu}_j=\mathbf{X}^{\mu}_j\boldsymbol\beta^{\mu}_j.\label{eq:mean_model_mu}
\end{IEEEeqnarray}
\subsubsection{Gaussian heteroskedastic}
The Gaussian heteroskedastic model features steady-state and short-term components for the log variance as defined in Eq \eqref{eq:noise_model}, and the same mean as in the homoskedastic model, 
\begin{IEEEeqnarray}{lLl}\label{eq:model_gauss_hetero}
    \boldsymbol{y}_j|\mathbf{X}^{\mu}_j,\mathbf{X}^{\sigma}_j\sim \mathcal{N}(\boldsymbol{\mu}_j,\boldsymbol{\sigma}^2_j).
\end{IEEEeqnarray}
We use again Gibbs sampling to simulate from the posterior distribution. Conjugate results apply for the mean regression parameters, but for $\boldsymbol\beta^{\sigma}_j$ we use the Newton approach detailed in section \ref{sec:gibbs_sampler} to sample from their conditional posteriors in a Metropolis step. The gradient and Hessian are with respect to a Gaussian likelihood, i.e.
\begin{IEEEeqnarray}{lLl}
    \mathbf{g}(\boldsymbol{\beta}_{\sigma})&=&\frac{\partial \ln p(\boldsymbol{\beta}_{\sigma}|-)}{\partial \boldsymbol{\beta}_{\sigma}}\nonumber\\
    &=&-\frac{1}{2}X_{\sigma}^T[\boldsymbol{1}-((\boldsymbol{y}-X_{\mu}\boldsymbol{\beta}_{\mu})^2\oslash\exp(X_{\sigma}\boldsymbol{\beta}_{\sigma}))]\nonumber\\
    &-&S_{\sigma}^{-1}(\boldsymbol{\beta}_{\sigma}-\boldsymbol{m}_{\sigma})
\end{IEEEeqnarray}
\begin{IEEEeqnarray}{lLl}
    \mathbf{H}(\boldsymbol{\beta}_{\sigma})&=&\frac{\partial^2 \ln p(\boldsymbol{\beta}_{\sigma}|-)}{\partial \boldsymbol{\beta}_{\sigma}\partial\boldsymbol{\beta}_{\sigma}^T}\nonumber\\
    &=&-\frac{1}{2}X_{\sigma}^T\left[(\boldsymbol{y}-X_{\mu}\boldsymbol{\beta}_{\mu})^2\oslash\exp(X_{\sigma}\boldsymbol{\beta}_{\sigma})\right]X_{\sigma}\nonumber\\
    &-&S_{\sigma}^{-1}.
\end{IEEEeqnarray}
\subsubsection{Student's $t$ homoskedastic}
As mentioned before, we use the scale-mixture representation of the Student's $t$ distribution, Eqs \eqref{eq:scale_mixture1}-\eqref{eq:scale_mixture2}, to facilitate Bayesian inference through Gibbs sampling. For the homoskedastic model, the mean is the same as in Eq \eqref{eq:mean_model_mu}, and the scale and degrees of freedom just scalars, i.e.
\begin{equation}\label{eq:model_t_homo}
    \boldsymbol{y}_j|\mathbf{X}^{\mu}_j\sim \mathcal{T}(\boldsymbol{\mu}_j,{\sigma}^2_j,{{\nu}_j}).
\end{equation}
The posterior distribution of the mean parameters is conjugate when conditioned on the augmented data $\boldsymbol{U}$, and $\alpha$. The posterior of the scale, $\sigma_j^2=\alpha^2\tau_j^2$, is also conjugate conditioned on the rest of model parameters. Full details on the sampler for this specific model are available at \cite{gelman2013bayesian}, which uses  a random-walk metropolis step on the logarithm of the degrees of freedom $\nu_j$. We use instead the Newton approach after reducing the model on Eq \eqref{eq:df_model} to just $\ln\nu = \beta_0$. 

\subsubsection{Student's $t$ heteroskedastic}
In this case, the likelihood and mean functions are the same as for the homoskedastic case, plus the model in Eq \eqref{eq:noise_model} for the log scale,
\begin{equation}\label{eq:model_t_hetero}
 \boldsymbol{y}_j|\mathbf{X}^{\mu}_j,\mathbf{X}^{\sigma}_j\sim \mathcal{T}(\boldsymbol{\mu}_j,\boldsymbol{\sigma}^2_j,{{\nu}_j}).\end{equation}
The Metropolis-within-Gibbs sampler is the one defined in section \ref{sec:gibbs_sampler}, keeping the degrees of freedom parameter $\nu_j$ as a scalar, and using the Newton proposals for both the scale and degrees of freedom.

\subsubsection{Student's $t$ full}
The last model includes linear regressions for all components, i.e. mean, scale and degrees of freedom as in Eqs  \eqref{eq:mean_model_mu}, \eqref{eq:noise_model}, and \eqref{eq:df_model}, respectively, 
\begin{equation}\label{eq:model_t_full}
 \boldsymbol{y}_j|\mathbf{X}^{\mu}_j,\mathbf{X}^{\sigma}_j,\mathbf{X}^{\nu}_j\sim\mathcal{T}(\boldsymbol{\mu}_j,\boldsymbol{\sigma}^2_j,{\boldsymbol{\nu}_j}).
\end{equation}
This is the general model defined in \ref{sec:general_model}, and the posterior inference is the same as for the heteroskedastic model, i.e. the Metropolis-within-Gibbs sampler defined in section \ref{sec:gibbs_sampler}, but in this case we have also a regression on the log $\boldsymbol{\nu}_j$, rather than a scalar.

\begin{table}[ht]
\centering
\caption{Covariate structure of the implemented models}
\label{tab:modelinfo}
\begin{tabular}{|cccccccc|}
\hline
Model & $Z^{\mu}$ & $Z^{\sigma}$ & $Z^{\nu}$ & $W^{\mu}$ & $W^{\sigma}$ & $W^{\nu}$ & Noise\\
\hline
\hline
Hist. average & \checkmark & - & - & - & - & - & $\mathcal{N}$\\
Random walk & - & - & -& - & - & - & $\mathcal{N}$\\
Gauss-Homosk. & \checkmark & - & -& \checkmark & - & - & $\mathcal{N}$\\
Gauss-Heterosk. & \checkmark & \checkmark & -& \checkmark & \checkmark & - & $\mathcal{N}$\\
$t$-Homosk. & \checkmark & - & -& \checkmark & - & - & $\mathcal{T}$\\
$t$-Heterosk. & \checkmark & \checkmark & -& \checkmark & \checkmark & - & $\mathcal{T}$\\
$t$-Full & \checkmark & \checkmark & \checkmark& \checkmark & \checkmark & \checkmark & $\mathcal{T}$\\
\hline
\end{tabular}
\end{table}

\section{Experimental setup}\label{sec:setup}
\subsection{Data description}

\begin{figure}[ht]
\centering
\includegraphics[width=3.5in]{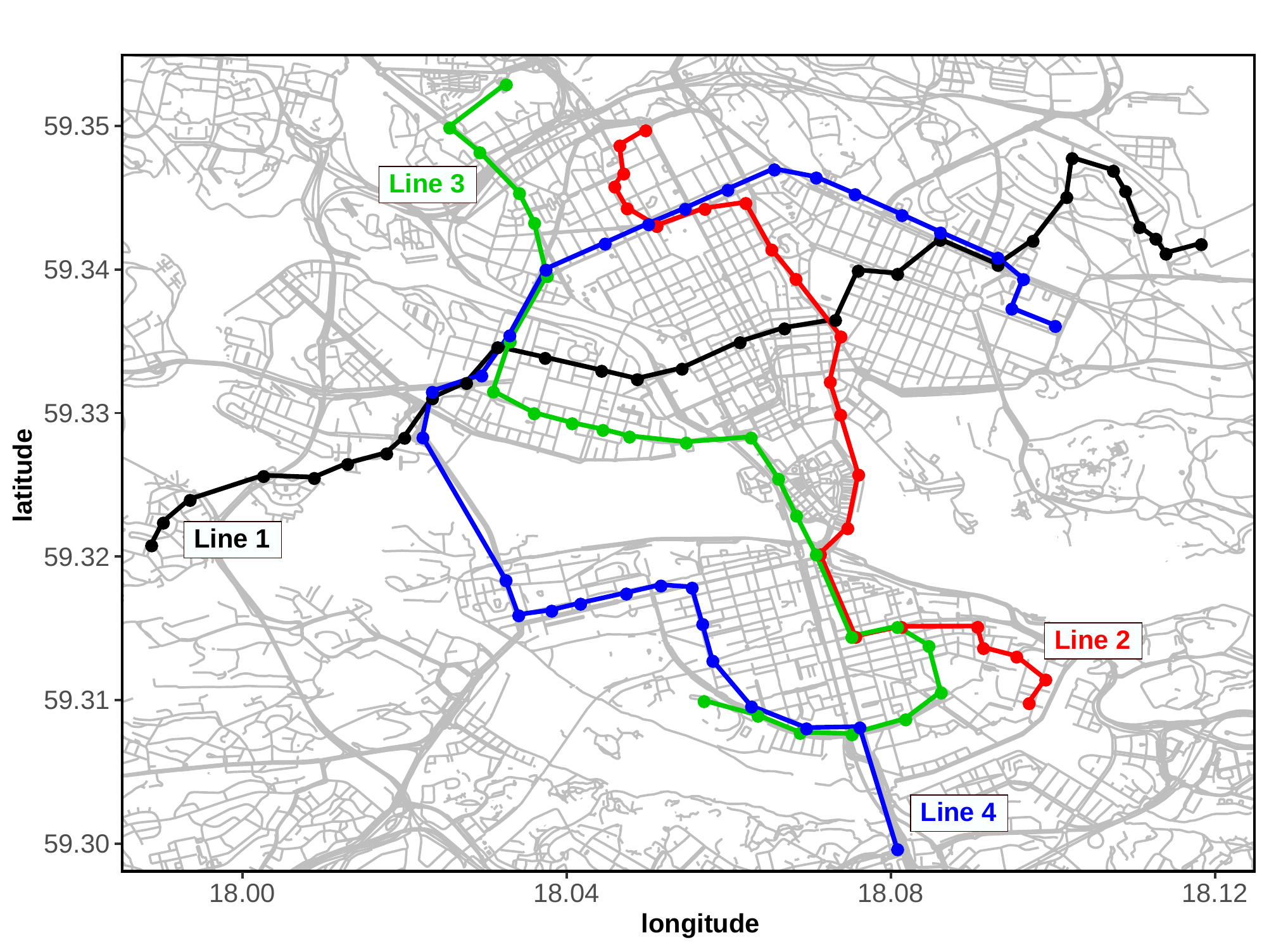}
\caption{Available high-frequency bus lines in central Stockholm.}
\label{fig_stockholm_map_bus}
\end{figure}

\begin{figure}[ht]
\centering
\includegraphics[width=3.5in]{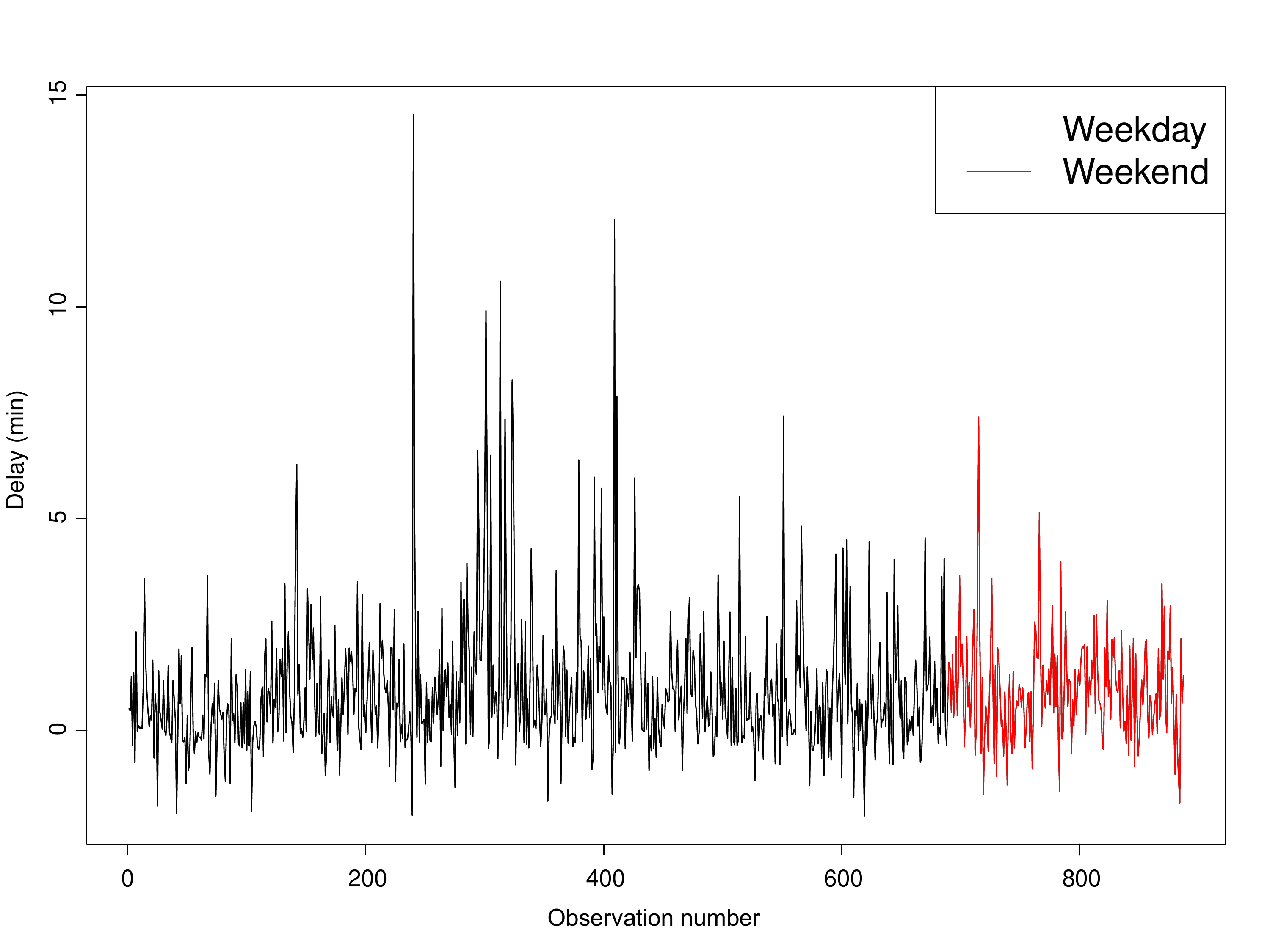}
\caption{Sample week of data for stop 8 at route 89 (week 10, 2017).}
\label{fig_sample_week_delay}
\end{figure}

We work with bus arrival times collected from GPS-equipped buses in
the city of Stockholm. The data consist on AVL records from four bus lines in the inner city during 2017, and was provided by Stockholm's public
transport agency (Storstockholms Lokaltrafik - SL). The geographical layout of the lines is displayed in Figure \ref{fig_stockholm_map_bus}. In total, the data comprises more than 15 million records containing bus and route information, timestamps, scheduled and actual arrival times to stops, distance covered, etc. Buses send data in an event-based fashion (e.g. regular stop, drive through, disturbances), but we keep the observations recorded when the bus is located at a stop, so we can calculate the actual delay for that specific stop and time using Eq \eqref{eq:delaydef}.
Table \ref{tab:routeintfo} below presents information on the number of stops, average headway during morning/evening rush hours and length of all available routes. These can be considered high-frequency routes as average headways are below 10 minutes in all cases \cite{kittelson2003transit}. The selected sample period is from January 1\textsuperscript{st} to June 30\textsuperscript{th} 2017, about 26 weeks, and we kept records from 6AM to 9PM, see Figure \ref{fig_sample_week_delay} for a sample of delay data for a specific week and stop. Route lengths range from 7 to 12 kilometers, with average headways between 5 and 7 minutes, and a total of 223 stops in the network. We selected the route 89 from Line 1 for the experiments. This route traverses the inner city from west to east and has the largest number of stops, 32. The data processing was coded in Python, and results in a set of stop-wise matrices containing irregular bus delay time-series (between 20,000-25,000 observations per stop) that can be fed to the models presented in the next section, which are implemented in R and C++. Source code for the model estimation is publicly accessible at \url{https://github.com/hecro459}.

\begin{table}[ht]
\centering
\caption{Available route's information}
\label{tab:routeintfo}
\begin{tabular}{|ccccc|}
\hline
Line & Route no. & Stops & Headway (min) & Length (Km)\\
\hline
\hline
1 & 86 & 31 & 6.04 & 9.4\\
1 & 89 & 32 & 6.07 & 9.6\\
2 & 2354 & 23 & 6.65 & 7.2\\
2 & 1093 & 24 & 6.73 & 7.3\\
3 & 3026 & 26 & 6.42 & 9.2\\
3 & 65 & 26 & 7.05 & 9.1\\
4 & 1075 & 31 & 5.95 & 11.6\\
4 & 190 & 30 & 5.86 & 11.4\\
\hline
\end{tabular}
\end{table}

\subsection{Features and model configuration }
In our experiments, a unique steady state feature matrix, $\mathbf{Z}$, is shared by the mean, scale and degrees of freedom models \eqref{eq:obs_model}-\eqref{eq:df_model}. We included 21 features to capture within-day hourly effects (15 dummies - 7AM to 9PM) and weekday patterns (6 dummies - Monday to Sunday). Some of the variables were removed to avoid multicollinearity, and in the weekday categories festivities were considered as Sundays. As remarked before, given an estimated model, the features in $\mathbf{Z}$ will determine the average delay in the absence of recent data, or at a distant prediction horizon. We do not focus on seasonal variability here, but it could be readily modeled in the steady-state as a function of e.g. week numbers.

The number of features (i.e. columns) in the short term matrices $\textbf{W}$ depend on the choice of $L$ and $P$. In our experiments, we found no significant improvement beyond $P=3$ previous data points, and $L=2$ preceding buses, so $L\times P=6$ features will be used, plus an intercept. Since the short term features are functions of past delays, entries of $\textbf{W}$ also depend on the current time $t$, and the geometric decay factor $\delta$, which we set to $\delta=0.96$ so previous data points will be shrunk by a factor of 0.5 and 0.9, after 20 minutes and one hour, respectively. A more refined model selection can be performed via a grid search of predictive performance on test data, Bayesian Optimization or similar techniques, but we keep the values above for simplicity.

In Section \ref{sec:methodology} short-term features for the model mean $\textbf{W}^{\mu}_j$ are defined as functions of past delays. On the other hand, elements in the feature matrices $\textbf{W}^{\sigma}_j,\textbf{W}^{\nu}_j$ for models \eqref{eq:noise_model},\eqref{eq:df_model}, correspond to the time-discounted absolute differences between pairs of delays from previous buses heading at a given stop $j$, i.e.  
 \begin{IEEEeqnarray}{lLl}
 	w_{jl}(t)&=&\lvert y_p(h_p(t,l))-y_{p-1}(h_{p-1}(t,l))\rvert\delta_j^{t-h_p(t,l)},
 \end{IEEEeqnarray}
for some $l=1,...,L$ and $p\in \mathcal{G}_j^*(t,l)$. Note that, in contrast with the steady state covariates $\mathbf{Z}$, each row vector $\mathbf{w}(t)$ in $\mathbf{W}$ is weighted with respect to the actual time $t_{\text{obs}}$ where a given delay $y_j(t)$ was observed. This means that we train our model with features $\mathbf{W}$ created from the previous delays that were available at the very moment the actual arrival time happened, i.e. $t=t_{\text{obs}}$ for both $y_j(t)$ and $\mathbf{w}(t)$. For out-of-sample predictions, however, we are rather interested in using features from $h$ minutes before the actual arrival time, that is, using $\mathbf{w}(t_{\text{obs}}-h)$ to predict $y(t_{\text{obs}})$. Therefore, short-run test features must be created for a given offset of $h$ minutes before $y(t_{\text{obs}})$ independently. These test datasets are useful for model benchmarking and evaluation: as the prediction horizon is increased, predictions are expected to converge to a historical average determined by $\mathbf{Z}$. We use a 80/20 split of the data, using (approximately) the first 20 weeks of the year 2017 for training, and weeks 21-25 for testing. 

The set of past delays and differences between delays from the current and previous buses are also expected to capture variability related to traffic conditions on the preceding stops in the route. Still, the actual running times between stops can easily be used as additional covariates in our regression model for unexpected traffic congestion \cite{yu2017using}\cite{achar2019bus}. If data on passenger counts is available, the number of boarding and/or alighting passengers at the preceding stations would also help modeling the fluctuations of passenger flows, which is another important short-term feature that affects bus arrival times \cite{zhou2017bus}\cite{tetreault2010estimating}.

Models are estimated using the corresponding Markov chain Monte Carlo (MCMC) approach, as specified previously, using crude maximum-likelihood estimates as starting values for $\boldsymbol{\beta}_{\mu},\boldsymbol{\beta}_{\sigma}$, and the intercept of $\boldsymbol{\beta}_{\nu}$. For the models that use the Newton's method, we take just two Newton iterations in the posterior sampling steps of both $\boldsymbol{\beta}_{\sigma},\boldsymbol{\beta}_{\nu}$. All MCMC algorithms are implemented in R, with inlined C++\cite{eddelbuettel2014armadillo} to accelerate the Newton iterations.

\section{Results and Discussion}\label{sec:results}
\subsection{Model benchmarking}\label{sec:results-bench}
The performance of the proposed models on real bus data is assessed by computing the log point-wise predictive density (LPPD) and mean absolute errors (MAE) on both training and test datasets. Let $y_i$ be the original $i$-th response and $\hat{y}_i$ its prediction. The MAE
\begin{IEEEeqnarray}{lLl}
	\sum_{i=1}^n\lvert y_i-\hat{y}_i\rvert/n
\end{IEEEeqnarray}
is a popular measure of performance for point prediction and has the advantage of being easy to calculate and interpret, but does not measure the accuracy of the full predictive distribution, which is the main object of interest here. For probabilistic predictions one should use measures that take into account the full predictive uncertainty, such as the LPPD for the test data
\begin{IEEEeqnarray}{lLl}
	\text{LPPD}&=&\sum_{i=1}^n\log\left(\sum_{s=1}^{S}p(y_i|\theta^s)/S  \right), 
\end{IEEEeqnarray}
where $\theta^s$ are draws from the posterior distribution $p(\boldsymbol{\theta}|\boldsymbol{y})$ based on the training data. We draw 20,000 samples from the posterior distribution of each model and discard 10,000 as burn-in. Since we use observations from a single network node (stop 8 in route 89), we are able to run the experiments on a regular PC with times ranging from 20 minutes for the random walk and the Gaussian homoskedastic model, to 3 hours for the $t$-Full model. 

\begin{table}[ht]
\centering
\caption{Model benchmarking on real delay data (MAE in seconds)}
\label{tab:modelbenchmark}
\begin{tabular}{|ccccc|}
\hline
Model & LPPD\textsubscript{train} & LPPD\textsubscript{test} & MAE\textsubscript{train} & MAE\textsubscript{test}\\
\hline
\hline
Hist. average & -85,429 & -22,607 & 61.79 & 75.55\\
Random walk & -82,637 & -22,150 & 33.48 & \textbf{41.08}\\
Gauss-Homosk. & -79,167 & -21,656 & 31.83 & 43.35\\
Gauss-Heterosk. & -77,801 & -21,183 & 31.01 & 42.10\\
$t$-Homosk. & -71,644 & -19,399 & 28.79 & 42.64\\
$t$-Heterosk. & -71,452 & -19,356 & \textbf{28.78} & 42.58\\
$t$-Full & \textbf{-71,380} & \textbf{-19,305} & 28.80 & 42.45\\
\hline
\end{tabular}
\end{table}

Table \ref{tab:modelbenchmark} present training and test LPPD and MAE for all models. As expected, performance on training data improves with model complexity, for both LPPD and MAE, with models featuring Student-$t$ noise achieving higher LPPD and lower errors in general. The steady-state prediction (historical average) yields the worse out-of-sample MAE, more than a minute delay, whereas the random walk stands as the best model for mean prediction. Both Gaussian and Student-$t$ models perform similarly in terms of mean errors. Turning to the more interesting LPPD measure, we see that the Gaussian models are dramatically outperformed by their robust counterparts, which are much better at modeling the distribution of the heavy-tailed delay data (see Fig.\ref{fig:hist_delay_residuals}). The $t$-Full is the winner among robust models, performing better than the homoskedastic and heteroskedastic alternatives in both training and test LPPD. It seems that the inclusion of relevant covariates on the scale and degrees-of-freedom spaces has paid off in terms of out-of-sample uncertainty prediction, which supports our preference for a very flexible model.  

\begin{figure}[ht]
\centering
\includegraphics[width=3.5in]{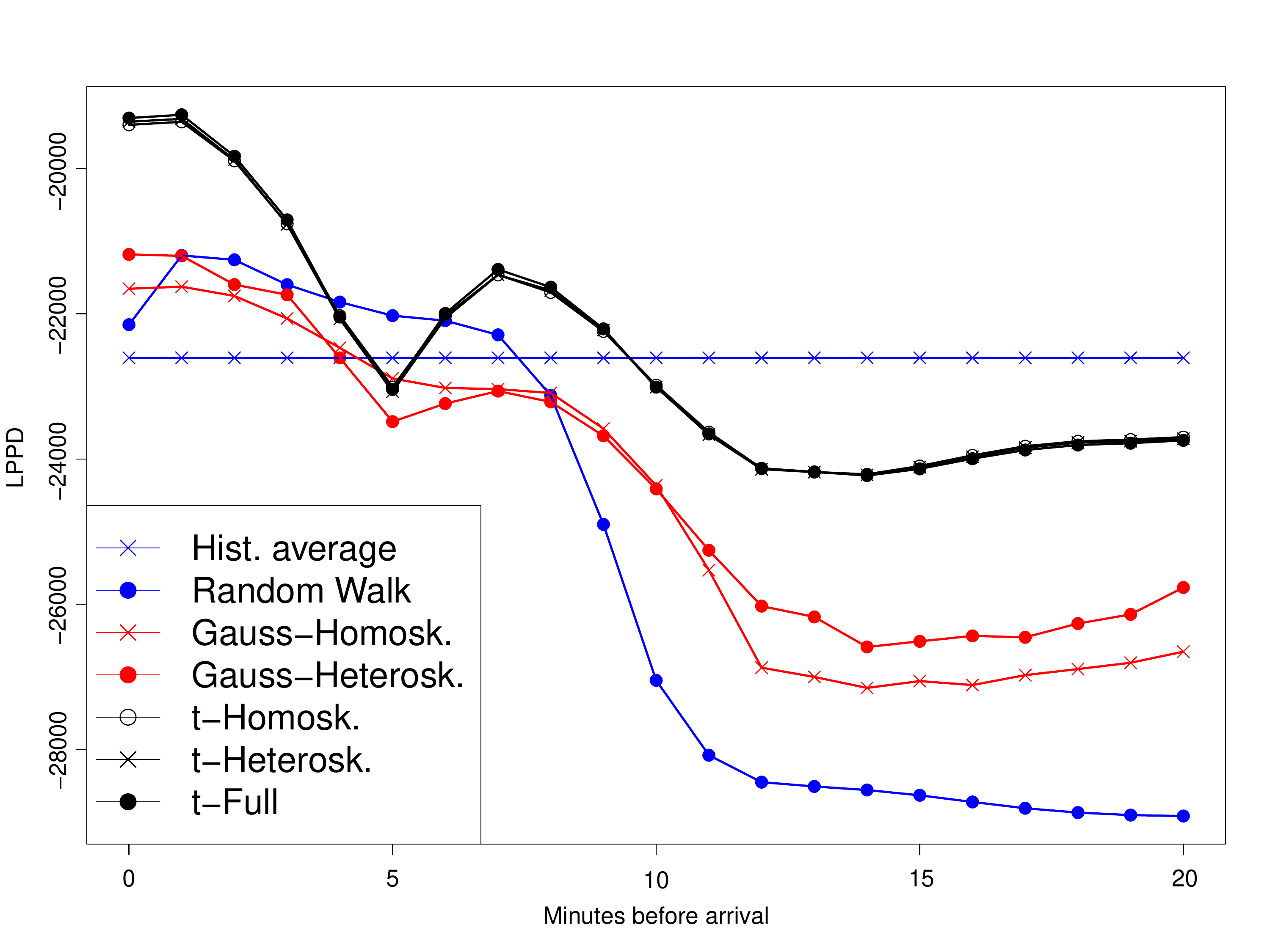}
\caption{Test LPPD for all models at different before-arrival times.}
\label{fig_lppd_curves}
\end{figure}

Figure \ref{fig_lppd_curves} shows LPPD curves that represent the model's probabilistic performance at different $h=\{0,1,\ldots,20\}$ minutes before the actual arrival time $t_{\text{obs}}$. On average, all robust models (black lines) outperform the Gaussian ones (red) up to 20 minutes before arrival, which is in line with the initial ($h=0$) results from Table \ref{tab:modelbenchmark}. The LPPD values for the $t$-Full model are consistently higher than those from the other two robust models, although it is difficult to appreciate directly from the plot. We can also see that the Gaussian heteroskedastic model outperforms the homoskedastic one in most cases. There is a gradual convergence to the steady-state by both Gaussian and Student-$t$ models, due to the lack of recent data points and the geometric decaying. Conversely, the performance of the random walk, which lacks a steady-state component, does not recover and worsens gradually as it moves away from the arrival time. The historical average would be the best choice in the long-run, as expected, and does a good job overall since it is using the important long-term predictors in $\mathbf{Z}^\mu$.
\subsection{Posterior analysis of the $t$-Full model}\label{sec:results-posterior}
\begin{table}[ht]
	\centering
	\caption{Posterior summaries for $\boldsymbol{\beta}_{\mu}$ - $t$-Full model on Stop 8, Route 89. Bold parameter name indicates HPD does not contain zero.}
	\label{tab:modelposterior}
	\begin{tabular}{|crrrrr|}
		\hline
		Parameters & Mean & Stdev & HPD\textsubscript{low} & HPD\textsubscript{upp} & IF\\
		\hline
		\hline
		\textbf{Intercept} &     38.43&      0.66&   37.37&   39.47&      9.42\\
		Hour\_8    &     -0.10&      0.69&   -1.19&    1.02&      4.34\\
		\textbf{Hour\_9}    &     -1.04&      0.61&   -1.97&   -0.03&      3.86\\
		Hour\_10   &     -0.01&      0.61&   -1.00&    0.93&      4.28\\
		\textbf{Hour\_11}   &      2.34&      0.63&    1.33&    3.33&      3.32\\
		\textbf{Hour\_12}   &      1.95&      0.61&    1.00&    2.94&      3.45\\
		Hour\_13   &      0.90&      0.61&   -0.13&    1.80&      3.41\\
		\textbf{Hour\_14}   &      1.27&      0.60&    0.34&    2.26&      5.24\\
		\textbf{Hour\_15}   &      2.70&      0.64&    1.71&    3.75&      5.10\\
		\textbf{Hour\_16}   &      5.32&      0.66&    4.25&    6.35&      3.63\\
		\textbf{Hour\_17}   &      5.32&      0.69&    4.15&    6.37&      4.96\\
		\textbf{Hour\_18}   &      2.34&      0.69&    1.21&    3.41&      6.07\\
		\textbf{Hour\_19}   &     -1.29&      0.73&   -2.45&   -0.13&     10.46\\
		\textbf{Hour\_20}   &     -2.93&      0.76&   -4.15&   -1.69&     10.51\\
		\textbf{Hour\_21}   &     -1.80&      0.77&   -3.04&   -0.58&      7.80\\
		Weekday\_2 &      1.06&      0.73&   -0.12&    2.22&      8.38\\
		\textbf{Weekday\_3} &      2.61&      0.75&    1.37&    3.74&      8.30\\
		Weekday\_4 &      0.34&      0.70&   -0.76&    1.48&      8.66\\
		Weekday\_5 &      1.03&      0.76&   -0.26&    2.16&      8.21\\
		\textbf{Weekday\_6} &    -10.76&      0.82&  -12.01&   -9.41&      8.51\\
		\textbf{Weekday\_7} &     -6.97&      0.88&   -8.35&   -5.54&     12.79\\
		\textbf{Delay\_l1\textsubscript{p1}}  &     83.72&      0.81&   82.49&   85.07&      5.16\\
		\textbf{Delay\_l1\textsubscript{p2}}  &     -6.76&      1.00&   -8.44&   -5.25&      4.89\\
		Delay\_l1\textsubscript{p3}  &     -0.76&      0.57&   -1.66&    0.15&      2.85\\
		\textbf{Delay\_l2\textsubscript{p1}}  &      6.05&      0.42&    5.38&    6.73&      4.30\\
		\textbf{Delay\_l2\textsubscript{p2}}  &     -1.88&      0.65&   -2.90&   -0.82&      4.00\\
		\textbf{Delay\_l2\textsubscript{p3}}  &     -2.61&      0.60&   -3.57&   -1.66&      4.13\\
		$R^2=0.63$ & & & & &\\
		\hline
	\end{tabular}
\end{table}

We focus on the output from the most elaborate model, $t$-Full, to evaluate the estimation and analyze interesting patterns. Table \ref{tab:modelposterior} presents summaries from the posterior distribution of the mean regression parameters, $\boldsymbol\beta_{\mu}=(\boldsymbol\alpha_{\mu}^\top,\boldsymbol\gamma_{\mu}^\top)^\top$, calculated from the 10,000 after-burn-in samples, including 90\% highest posterior density intervals (HPD). The table also includes the inefficiency factor (IF) for each parameter, a common measure of numerical efficiency of the MCMC results, which is defined as
\begin{IEEEeqnarray}{lLl}
	\text{IF}(\beta)&=&1+2\sum_{j=1}^\infty{\rho}_j(\beta),
\end{IEEEeqnarray}
where ${\rho}_j(\beta)$ is the sample autocorrelation at lag $j$ for the chain of parameter $\beta$. The value of the Bayesian $R^2$ is 0.63, which is calculated as the posterior median of the indicator proposed by \cite{gelman2019r}. Some within-day delay patterns have been captured by the parameters corresponding to the hourly dummy variables. It seems that the model have identified increased congestion at the evening rush hour (16:00-18:00), but not during the morning one. As expected, there is a sudden delay reduction after 19:00, and a plateau during the lunch hours, although the latter effects do not seem to be that significant as some of their HPD intervals include zero. Weekends present reduced delay levels, particularly during Saturdays, and Wednesdays stand out with the stronger positive effect on delays. No other noticeable pattern is appreciated at weekday level. The last six parameters correspond to the spatiotemporal effects related to available delay information on the incoming ($l=1$) and the immediately previous buses ($l=2$), on the preceding stops (i.e. $8,7,\ldots $). Note that each of the $P=3$  
effects per bus do not refer to minute-wise lags, but the three most recent delay observations. Overall, it seems that for this specific stop the most recent delay data points from both buses are enough for the mean prediction, as the effects of  Delay\_l1\textsubscript{p2-3} and Delay\_l2\textsubscript{p2-3} are considerably lower than for the most recent observation in all cases. The stronger effect is at the parameter Delay\_l1\textsubscript{p1}, corresponding to the most recent delay recorded from the incoming bus at a given time. This makes sense since, aside from a bus-bunching situation or a severe delay, this will correspond to a record from a bus at a nearby stop. The estimate for the second most recent delay of the incoming bus, Delay\_l1\textsubscript{p2}, is also significant, but with a comparatively small effect. With respect to MCMC convergence, the expected sample sizes and inefficiency factors indicate a very efficient sampling for all the model parameters. The acceptance rates in the Newton-based Metropolis-Hastings steps are about 58\% and 55\% for $\boldsymbol\beta_{\sigma}$ and $\boldsymbol\beta_{\nu}$, respectively.

\begin{figure}[ht]
\centering
\includegraphics[width=3.5in]{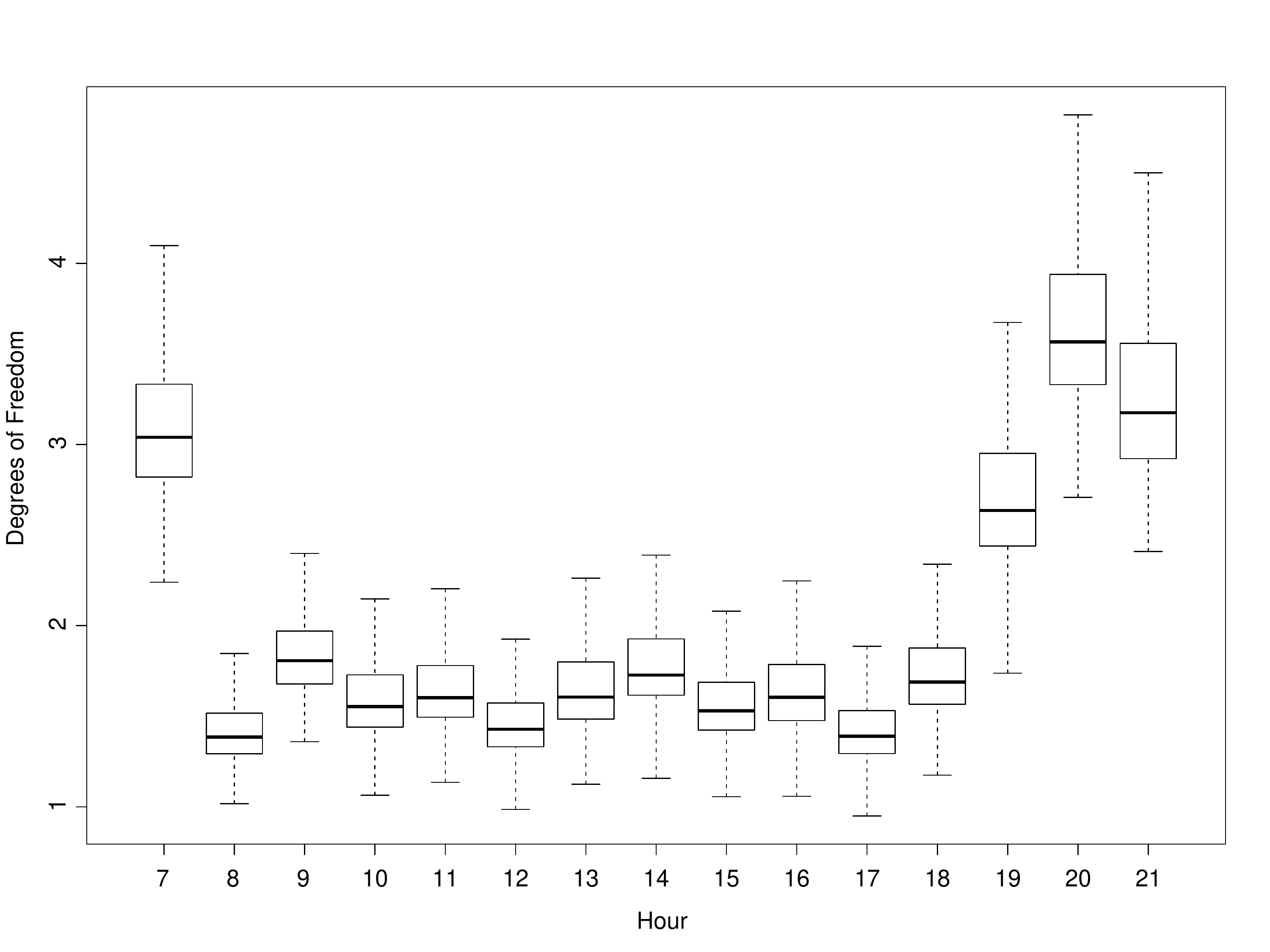}
\caption{Estimated degrees of freedom $\hat{\nu}$ per hour of the day on training data.}
\label{fig_estimateshour}
\end{figure}

Analogous estimates $\hat{\boldsymbol\beta}_{\sigma},\hat{\boldsymbol\beta}_{\nu}$ are available for the regressions on the log scale and degrees of freedom though we do not present tables for the sake of brevity, but describe them generally. Weekday parameters show an increase in degrees of freedom during weekends, particularly on Sunday, which indicate a reduced occurrence of extreme events. The same applies in the early morning hours, and in the late evening, also for the scale. Figure \ref{fig_estimateshour} shows the fitted degrees of freedom per hour of the day for the training data. Delay data seems to be consistently heavy-tailed as the model estimates $\hat{\nu}<10$ in all cases, which indicates a presence of outlying observations regardless of the hour of the day. Still, we see how the regression model on the $\log\nu$ space yields a clear pattern related to traffic congestion within the day, with an increased probability of extreme delays between 8AM and 6PM. Note that many estimated values of $\hat{\nu}$ are smaller than $2$, and that for the $\mathcal{T}(\mu,\sigma^2,\nu)$ distribution the variance is not defined for $\nu<2$, which makes difficult to interpret the estimated variance. For the estimates corresponding to the absolute differences between recent observations, we observe significant effects for the incoming and immediately previous buses, both in scale and degrees of freedom regressions. The stronger effect for absolute differences in $\log\sigma^2$ is for the most recent pair of observations from the incoming bus, and has negative sign; the same applies to $\log\nu$. This indicates that the model detects a correlation between recent high variability between delays of previous buses, with an increased probability of a tail event in the current one, and a lower scale. These effects, however, seem comparatively small in magnitude, and other potential explanatory variables should be investigated, in order to capture short term variations in scale and degrees of freedom more effectively. 

\begin{figure}[ht]
\centering
\includegraphics[width=3.5in]{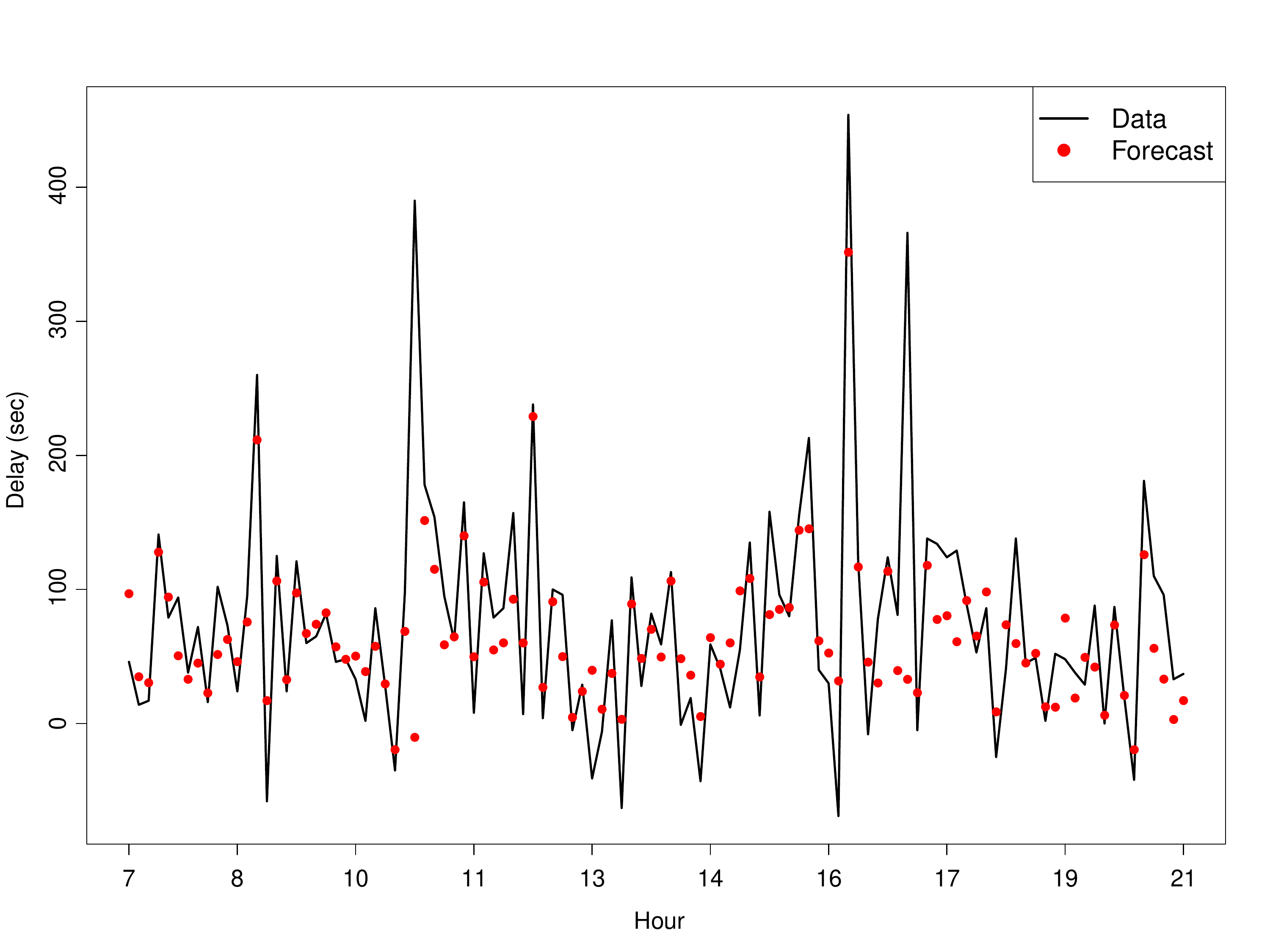}
\caption{Four-minutes ahead delay forecasts for a sample day, test data.}
\label{fig_fit_mean}
\end{figure}

Finally, we show in Figure \ref{fig_fit_mean} four-minutes-ahead mean forecasts on a selected day from the test dataset. The model seems to reasonably capture the overall delay distribution throughout the day, and to capture many of the outlying observations, although some of the largest delays ($>5$ min) are underestimated. The MCMC efficiency for estimating $\boldsymbol\beta_{\sigma},\boldsymbol\beta_{\nu}$  is worse than for the mean model, with average inefficiency factors around $30$ for most parameters. This is expected, given 
the difficulty of the inference on $\log\sigma^2$ and $\log\nu$ spaces. 

\subsection{Uncertainty forecasting}\label{sec:results-uncert}
\begin{figure}[ht]
\centering
\includegraphics[width=3.5in]{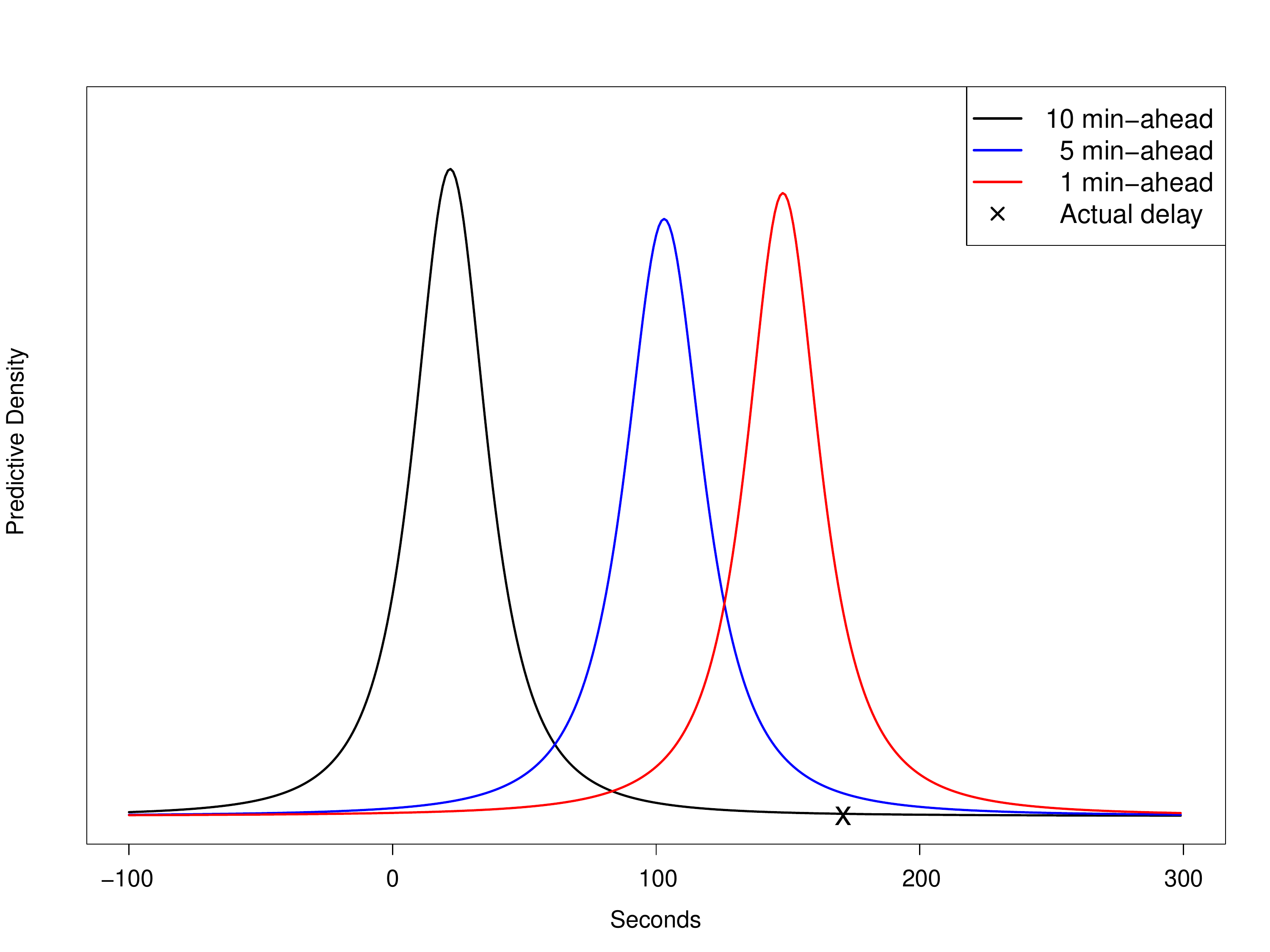}
\caption{Posterior predictive densities at different minutes-ahead for a test data point on a selected Wednesday. }
\label{fig_uncertainty_pred}
\end{figure}
Bayesian methods provide probabilistic forecasts through the posterior predictive distribution, which takes into account the full uncertainty about the parameters $\theta$ when predicting new data observations $y^{\text{new}}$ for given test covariates $\textbf{x}^{\text{new}}$ and the training data $\textbf{y},\textbf{X}$. In our case, the posterior predictive can not be derived in closed-form, so we approximate it using $S$ posterior MCMC samples as
\begin{IEEEeqnarray}{lLl}
	p(y^{\text{new}}| \textbf{x}^{\text{new}},\textbf{y},\textbf{X})&=&\int_{\theta}p(y^{\text{new}}|\textbf{x}^{\text{new}},\theta)p(\theta|\textbf{y},\textbf{X})d\theta\\
	&\approx&\sum_{s=1}^{S}p(y^{\text{new}}|\textbf{x}^{\text{new}},\theta^s)/S.  
\end{IEEEeqnarray}

In Figure \ref{fig_uncertainty_pred} we illustrate probabilistic forecasts from the $t$-Full model for a given snapshot of the data. The test point, marked with a black cross, represents a single outlying observation during the afternoon (around 2:30PM) on a selected Wednesday. The actual bus delay is of 171 seconds, about 3 minutes. We created corresponding predictive distributions at different times before the actual arrival time and computed HPD intervals. We see how predictions are initially more conservative and then move gradually to the actual delay. For 10 and 5 minutes in advance, the $95\%$ prediction intervals are of $[-37,114]$ and $[30,200]$ seconds, respectively. The predictive density one minute before the actual arrival time is very close to the actual delay, as the most recent observation on the incoming bus (possibly generated when arrives to the previous stop) is now available to the model, with a predictive median of $148$ seconds. The adequacy of the generated intervals can be assessed quantitatively in terms of e.g. length and coverage, see \cite{khosravi2011prediction}\cite{pereira2014metamodel} for applications to bus travel time and traffic prediction. Probabilistic modeling also allows to answer questions like e.g. \emph{"What is the probability that the bus is at least one minute late?"}. Tail probabilities such as $P(y^{\text{new}}\ge60)$ can be readily calculated from the predictive densities. The 10 minute-ahead forecast, which relies more on the steady-state and discounted recent observations, estimates a conservative $P(y^{\text{new}}\ge60)=0.02$. On the other hand, 5 and 1 minute-ahead forecasts yield $93\%$ and $98\%$ respective chances that next bus is coming at least one minute late. 

We can also contrast the ability of different models to capture an extreme event. Now we select another test point corresponding to a delay of 489 seconds, about 8 minutes, for which the models fail to get close in posterior mean. In this case, tail probabilities become too small for a direct inspection, but we calculate odd ratios for a comparative analysis. For instance, the odds ratio of predicting a delay greater than three minutes for the $t$-Full model against the Gaussian heteroskedastic is calculated as
\begin{IEEEeqnarray}{lLl}
	\text{OR}=\frac{p_\mathcal{T}(y^{\text{new}}\ge180)/p_\mathcal{T}(y^{\text{new}}<180)}{p_\mathcal{N}(y^{\text{new}}\ge180)/p_\mathcal{N}(y^{\text{new}}<180)},
\end{IEEEeqnarray}
where $p_\mathcal{T}$, $p_\mathcal{N}$ are posterior predictive densities for the robust and Gaussian models. The odds ratios for delays larger than 3, 5 and 7 minutes are of $13.46$, $14.29$ and $12.93$, respectively. These results show that, even though both alternatives fall short in mean prediction, the predictive density for the model with Student-$t$ errors is able to place a (comparatively) larger mass of probability towards the outlying observation. 

\subsection{Spatial analysis}
\begin{figure}[hb]
	\centering
	\includegraphics[width=3.5in]{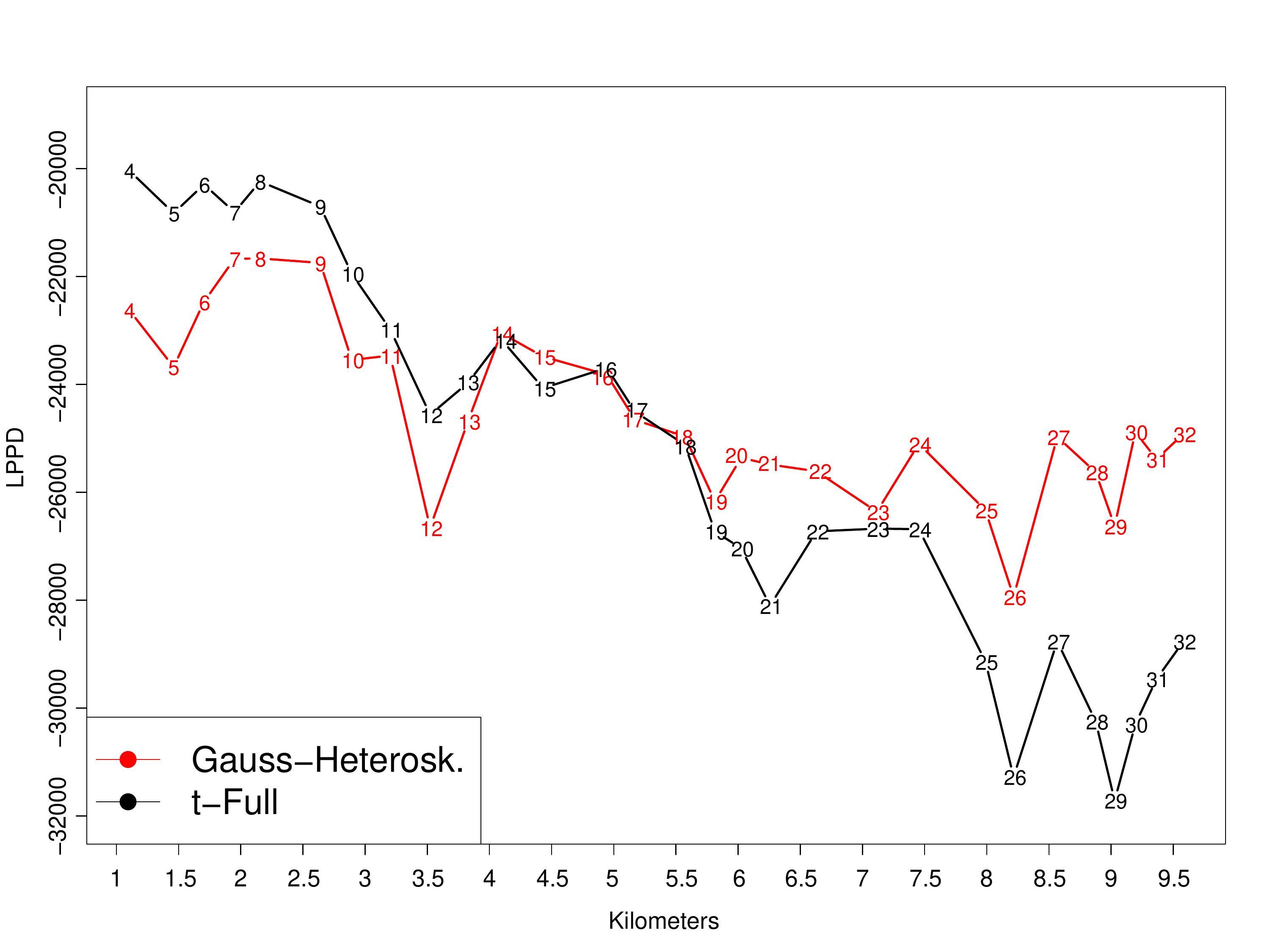}
	\caption{Spatial out-of-sample performance (test LPPD) for top models through route 89. The stop numbers are displayed at each data point.}
	\label{fig_spatial}
\end{figure}

We conclude the results with a brief discussion on spatial aspects from a route-level perspective. Figure \ref{fig_spatial} presents test LPPD's for the best Gaussian and Student-$t$ models through the route under study (stops 4 to 32, about 9 kilometers). Each curve represent the average LPPD for the  $h=\{0,1,\ldots,5\}$ minutes-before-arrival-time predictions for the respective model. In general, the predictive power of both alternatives degrades as buses progress along the route, with the $t$-Full model being outperformed by the Gaussian heteroskedastic at the end of the trip. It seems that the robust model is more adequate for predictions at the beginning of the route, up to stop 13, and the Gaussian at the end, after stop 20. The inspection of the moments for the delay distributions of all stops within the route will help in the interpretation of these results.

\begin{figure}[ht]
	\centering
	\includegraphics[width=3.5in]{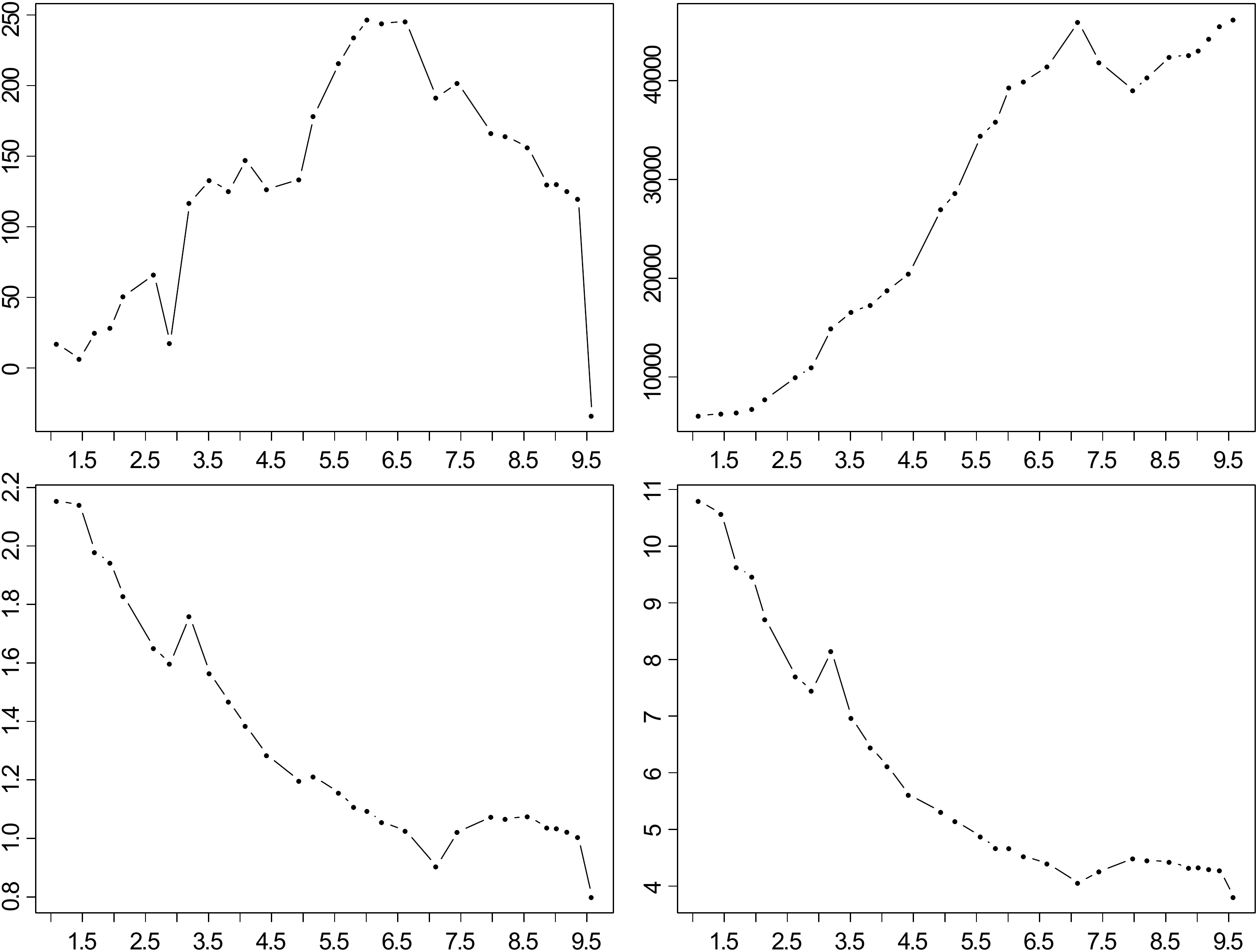}
	\caption{Moments of the delay distributions for stops in Route 89: mean (top-left), variance (top-right), skewness (bottom-left), and kurtosis (bottom-right). The $x$-axis represent route length in kilometers. }
	\label{fig_mom}
\end{figure}

The first four sample central moments (i.e. mean, variance, skewness and kurtosis) for the delay distributions at each stop are presented in Figure \ref{fig_mom}. The mean delay across the route has an inverted U-shape. Average delays increase steadily up to 4 minutes in station 22 (out of 32), and then decrease at the end of the route, perhaps due to a combination of traffic (peripheral streets), passenger (less boardings) and driver (catching-up) factors. The highest increase in mean delay happens after Fridhemsplan (stop 10) and H\"otorget (stop 16) stations, two popular hubs in central Stockholm. Variance, on the other hand, increases throughout the trip, with the steepest increase right after Stockholm's major transport hub, Cityterminalen (stop 15). A higher variability in passenger boardings, and perhaps bus-bunching may explain the rise in variability around the central hubs of the network. The kurtosis decreases throughout the route, representing a gradual convergence toward Gaussianity. This explains the switching pattern from Figure \ref{fig_spatial}: as the delay distributions become less thick-tailed at the end of the route, the robust Student-$t$ model is outperformed by the heteroskedastic Gaussian one. We see from these results that extreme delays in the tails of the distribution are less likely to happen at the end of this bus route, even though variability steadily increases. This is an interesting outcome, as one would expect that buses operating in inner-city high-frequency services can not catch-up easily if they fall behind schedule due to the short headways. 

The results in Figures \ref{fig_spatial} and \ref{fig_mom} illustrate the great complexity of arrival time prediction in an urban scenario, even within the same bus route, with the distributional properties of the variables of interest varying over space. Note that the Gaussian model is a special case of the full Student-$t$ model when the degrees of freedom parameter has no features except a large intercept. The results in Figures \ref{fig_spatial} and \ref{fig_mom} therefore suggest that $t$-Full model is over-parametrized for the stops at the end of the route, and that regularization of the regression coefficients in the degrees of freedom could help the model to recover the Gaussian distribution at those stops. One particularly attractive form of regularization for our model is Bayesian variable selection, which can be easily implemented by jointly sampling binary variable selection indicators and the regression coefficients in the finite-step Newton proposal step; see \cite{li2010flexible} for details. 

We stress that the previous results, though interesting, are aggregated for all time intervals (and weekdays, weeks, etc.), and a proper fine-grained spatiotemporal analysis must be carried out to avoid reaching premature conclusions. Aside from disaggregating the results according to the steady-state temporal covariates, as suggested above, increasing the number of routes (and therefore stops) and geographical coverage would also enhance the spatial analysis from a network perspective, which may reveal complex dependencies in terms of the model's performance. In that case, it would be also possible to contrast the results from both low and high-frequency services, which cannot be currently done with the data available in this study.

\section{Conclusion}\label{sec:conclusion}
Providing transport users and operators with accurate forecasts on travel times is challenging due to a highly stochastic traffic environment. In this paper we develop a robust model for real-time bus travel time prediction that depart from Gaussianity by assuming Student-$t$ errors. The proposed approach considers spatiotemporal features from the route and previous bus trips, and allows for a flexible regression modeling on mean, variance and degrees-of-freedom spaces. We provide an efficient Metropolis-within-Gibbs algorithms for Bayesian inference. Experiments are performed using real data from high-frequency urban buses in Stockholm, Sweden. 

Results show that Student-$t$ models outperform Gaussian ones in terms of log-posterior predictive density to forecast bus delays at specific stops. Even though there is little variation in mean predictions, which are below 1 minute in most cases, our results illustrate the importance of accounting for predictive uncertainty in model selection. Importantly, the flexibility induced by the inclusion of covariates on the scale and degrees-of-freedom spaces has clearly paid off in terms of out-of-sample predictive performance. Estimates from linear regressions in the most elaborate model capture various patterns in the delay distribution, such as rush/non-rush hours and weekday/weekend peaks, etc. For mean delay prediction, the stronger spatiotemporal effects are relative to incoming buses from immediately previous stops, which is in line with many recently developed models. The estimated regression on the degrees-of-freedom reveals a higher probability of extreme observations during working hours. Bayesian inference naturally allows for generating predictive intervals, and calculating the probability of e.g. incoming buses coming late, by using the posterior predictive distributions.

The results are promising, but extensive simulations must be carried out with larger databases to obtain more consistent evidence, investigate wider spatiotemporal network effects and contrast results between high and low-frequency services. Adding additional features such as bus and segment characteristic may also help predictions, as well as bus running times and passenger counts, which are important short-term variables affecting arrival times. From the methodological standpoint, the noninformative priors used here can be straightforwardly replaced by regularization priors to prevent overfitting. Variational inference could be also an alternative to MCMC, in order to scale up to massive datasets.

%
\section*{Acknowledgment}
This work was partially supported by the Wallenberg AI, Autonomous Systems and Software Program (WASP) funded by the Knut and Alice Wallenberg Foundation, and by the Swedish Research Council (Grant 2020-02846). We thank SL (Storstockholms Lokaltrafik) for providing the bus data.




%



\bibliography{bibtex/bib/references.bib}
\bibliographystyle{IEEEtran}
%
\begin{IEEEbiography}[{\includegraphics[width=1in,height=1.25in,clip,keepaspectratio]{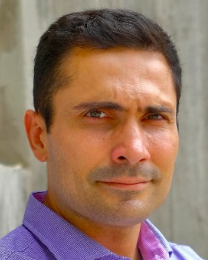}}]{Hector Rodriguez-Deniz} received the B.Eng. degree in Computer Engineering from the University of Las Palmas de Gran Canaria (Spain), and the master's in Statistics from Link\"oping University (Sweden). He is currently a doctoral student in Statistics at Link\"oping University within the Wallenberg AI, Autonomous Systems and Software program (WASP). His research interests include spatiotemporal probabilistic models, transportation networks, and machine learning. 
\end{IEEEbiography}

\begin{IEEEbiography}[{\includegraphics[width=1in,height=1.25in,clip,keepaspectratio]{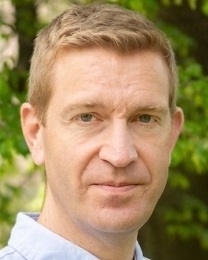}}]{Mattias Villani}
received the Ph.D. in Statistics from Stockholm University. He is currently Professor of Statistics at both Stockholm University and Link\"oping University in Sweden. His current research interests include Bayesian methodology for large-scale problems, in particular spatiotemporal modeling with applications in transportation, neuroimaging and economics. \end{IEEEbiography}
\vfill





\end{document}